# Electrically switchable one-dimensional quadrupolar excitons in lateral double heterojunctions

Ryo Kitaura*
*Research Center for Materials Nanoarchitectonics, National Institute for Materials Science, 1-1 Namiki, Tsukuba, Ibaraki 305-0044, Japan and Graduate School of Chemical Sciences and Engineering, Hokkaido University, 5, Kita 8 Nishi, Kita-ku, Sapporo, Hokkaido 060-8628, Japan*

Two neighboring lateral interfaces provide a spatial degree of freedom for controlling one-dimensional charge-transfer excitons within a single semiconductor monolayer. We investigate a type-II $WS_2$-$MoS_2$-$WS_2$ double heterojunction using an effective-mass two-particle Hamiltonian with a screened Coulomb interaction. For equivalent left and right interfaces at zero electric field, inter-interface coupling produces energetically split even- and odd-parity exciton states, each with zero permanent dipole. An electric field perpendicular to the interfaces continuously converts the lower state from a quadrupolar superposition with a quadratic Stark shift into a predominantly single-interface dipolar exciton with an approximately linear shift. The spatially resolved calculation gives binding energies of approximately 104 and 100 meV, a doublet splitting of 4.6 meV, and a crossover field of 0.87 V/$\mu$m for a representative 1.5-nm $MoS_2$ strip. Projection tests show that the lowest doublet controls the response near this crossover. Strip width tunes the coupling much more strongly than the binding energy, providing geometric control of the low-field Stark sensitivity. These results establish a continuum-model route to electrically reconfigurable one-dimensional quadrupolar excitons.

## I. INTRODUCTION

Lateral heterostructures consist of different monolayer semiconductors joined side by side through covalent bonds [1, 2]. Epitaxial growth has realized atomically sharp junctions [3], sequential multi-junction structures [4], and coherent lateral superlattices [5]. In these two-dimensional semiconductors, strong electron-hole attraction makes excitons central to the optical response [6, 7]. Lateral heterostructures can host various excitonic species, including excitons within either material and excitons bound across the interface [8, 9]. In particular, at a type-II junction the electron and hole favor opposite materials but can remain Coulomb-bound as a charge-transfer (CT) exciton, confined across the junction and free to propagate along it. Single-interface theory has established that this spatial separation is compatible with appreciable binding and a large in-plane dipole [8, 9]. Experiments have identified interface-related optical resonances [9] and a local reduction of photoluminescence intensity near the junction [10]. A second, nearby interface introduces an additional question: can two oppositely oriented CT configurations form a coherent exciton whose electric response differs from that of either interface alone?

Quadrupolar excitons in vertical heterostructures were predicted theoretically [11] and subsequently observed in tunnel-coupled heterotrilayers [12–14] and bilayers [15]. They form through hybridization of interlayer excitons with opposite dipoles. For equivalent outer layers, the hybridized states have no permanent dipole at zero electric field and exhibit a quadratic low-field Stark response.

* KITAURA.Ryo@nims.go.jp

Lateral interfaces offer a route to realize such multipolar states in one dimension. For a chosen vertical stack, hybridization is set by its interlayer tunneling and stacking configuration. A lateral double junction adds a directly designable in-plane length: the strip width sets the separation of the interfaces and thereby tunes their coupling. Atomically sharp $MoS_2$-$WS_2$ structures with nanometer and subnanometer strip widths provide an experimental basis for this geometric control [16]. An in-plane field can then select between the two parallel exciton channels. Together, strip-width design and electric bias offer complementary controls of a tunable one-dimensional quadrupolar exciton.

We consider a $MoS_2$ strip embedded in $WS_2$, forming two parallel interfaces as illustrated in Fig. 1(a). The electron preferentially occupies the central region, whereas the hole can lie on either side. Earlier theory established bound excitons in lateral double junctions and showed how their carrier distributions depend on strip width [8]. Here, we examine the coherent coupling of the opposite-dipole configurations and their electric-field-driven conversion. For equivalent left and right interfaces at zero electric field, coupling between the left- and right-interface configurations $|L\rangle$ and $|R\rangle$ produces a lower even-parity state and an upper odd-parity state, separated by an energy gap. Each has zero permanent dipole and a nonzero electric quadrupole moment, evaluated from the electron and hole charge distributions in Sec. S16 of the Supplemental Material [17]. The coherent, opposite-dipole superposition constitutes the quadrupolar configuration; a transverse field continuously biases it toward a single-interface dipolar configuration [Fig. 1($c$)].

We solve a spatially resolved two-particle effective-mass model that treats carrier confinement, screened Coulomb attraction, and inter-interface overlap together.

The lowest coupled charge-transfer exciton pair (CT1) supports a quadrupolar-to-dipolar crossover accompanied by a redistribution of optical transition strength. Its binding energy is much larger than the hybridization scale, allowing the switching field to be tuned strongly through strip width while retaining substantial binding. Comparison with the full finite-field calculation establishes when the lowest pair describes the energy and dipole response and where higher exciton states redistribute optical transition strength. Together, these results connect one-dimensional confinement, multipolar character, and lateral electrical control.

## II. CONTINUUM MODEL

### A. Two-particle Hamiltonian

The interfaces extend along $y$, with $x$ perpendicular to them [Fig. 1(a)]. For an electron and a hole at ($x_\mathrm{e}$, $y_\mathrm{e}$) and ($x_\mathrm{h}$, $y_\mathrm{h}$), we define the center-of-mass coordinates $X = (m_\mathrm{e}x_\mathrm{e} + m_\mathrm{h}x_\mathrm{h})/M$ and $Y = (m_\mathrm{e}y_\mathrm{e} + m_\mathrm{h}y_\mathrm{h})/M$, and the relative displacement $r = (x,\, y) = (x_\mathrm{e}$ - $x_\mathrm{h},\, y_\mathrm{e}$ - $y_\mathrm{h})$. Here $m_\mathrm{e}$ and $m_\mathrm{h}$ are the effective masses, $M = m_\mathrm{e} + m_\mathrm{h}$ is their sum, and $\mu = m_\mathrm{e}m_\mathrm{h}/M$ is the reduced mass. Because the potentials are independent of $Y$, the conserved longitudinal center-of-mass wave vector $K_y = (p_{e,y} + p_{h,y})/\hbar$ labels a separable plane-wave factor $\exp(iK_y\mathrm{Y})$. The quantities $p_{e,y}$ and $p_{h,y}$ are the electron and hole envelope momenta parallel to the interfaces. This motion contributes $\hbar^2K_y^2/(2M)$ to the energy. We set $K_y = 0$ and solve for the remaining envelope $\Psi(X,\, x,\, y)$, retaining the relative motion along $y$. The transverse center-of-mass coordinate $X$ remains coupled to the relative coordinates by the interface potentials. With $x_\mathrm{e} = X + (m_\mathrm{h}/M)x$ and $x_\mathrm{h} = X$ - $(m_\mathrm{e}/M)x$, the reduced Hamiltonian $\hat{H}_\mathrm{red}$ acting on $\Psi(X,\, x,\, y)$ is

$$\begin{aligned}\hat{H}_\mathrm{red} = &-\frac{\hbar^2}{2M}\partial_X^2 - \frac{\hbar^2}{2\mu}\nabla_{\boldsymbol{r}}^2 + V(r) + eFx \\ &+ V_\mathrm{e}\Big(X + \frac{m_\mathrm{h}}{M}x\Big) \\ &+ V_\mathrm{h}\Big(X - \frac{m_\mathrm{e}}{M}x\Big),\end{aligned} \tag{1}$$

The subscript red denotes removal of the free longitudinal center-of-mass motion only; the transverse coordinate $X$ is retained. Hats denote operators, and Eq. (1) gives their coordinate representation. The electric field $F$ is an external parameter, not a coordinate argument of the wavefunction.

Here $\hbar$ is the reduced Planck constant, $e$ is the positive elementary charge, and $F$ is the electric field in the positive $x$ direction. The interaction $V$ depends on the electron-hole distance $r$. The band-offset potentials $V_\mathrm{e}$ and $V_\mathrm{h}$ depend on both $X$ and $x$, as shown explicitly in Eq. (1). Translating the pair perpendicular to the interfaces changes these potentials even at fixed electron-hole separation. Thus separation of the kinetic terms does not separate the transverse center-of-mass motion from the internal motion. The masses are taken to be constant across the structure.

We define the carrier confinement energies relative to the $MoS_2$ conduction edge and the $WS_2$ valence edge: $V_\mathrm{e}(x) = E_\mathrm{c}(x)$ - $E_\mathrm{c}^{\mathrm{MoS_2}}$ and $V_\mathrm{h}(x) = E_\mathrm{v}^{\mathrm{WS_2}}$ - $E_\mathrm{v}(x)$. The reversed order for $V_\mathrm{h}$ reflects the energy cost of a hole, rather than a valence-band electron. The positive band offsets are $\Delta\ E_\mathrm{c} = E_\mathrm{c}^{\mathrm{WS_2}}$ - $E_\mathrm{c}^{\mathrm{MoS_2}}$ and $\Delta\ E_\mathrm{v} = E_\mathrm{v}^{\mathrm{WS_2}}$ - $E_\mathrm{v}^{\mathrm{MoS_2}}$. To express the band-edge profile in Fig. 1(b), we introduce the smooth central-region function

$$s(x) = \frac{1}{2}\left[\tanh\frac{4(x+L/2)}{w} - \tanh\frac{4(x-L/2)}{w}\right], \tag{2}$$

The confinement potentials defined above then take the form

$$\begin{aligned}V_\mathrm{e}(x) &= \Delta E_\mathrm{c}[1 - s(x)], \\ V_\mathrm{h}(x) &= \Delta E_\mathrm{v} s(x),\end{aligned} \tag{3}$$

$L$ is the central-strip width and $w$ controls the smooth transition at each interface. Following the hyperbolic-tangent profiles used for lateral interface excitons [8, 9], Eq. (2) combines two steps with the $\tanh(4x/w)$ convention of Ref. [9]. We choose $w = 1$ nm, corresponding to a 10%-90% transition distance of 0.549 nm (Sec. S2 [17]). Atomically sharp chemical interfaces and nanometer-scale strips have been observed [16]. Their chemical sharpness does not itself determine the electronic band-edge transition width; the 1.6-nm length reported in Ref. [16] instead describes wavefunction penetration.

We use the Rytova-Keldysh interaction, widely employed in effective-mass calculations of two-dimensional excitons and trions [20–22] and in modeling exciton Rydberg spectra [23], with a spatially uniform screening parameter,

$$\begin{aligned}V(r) &= -\frac{\pi C}{2\kappa r_*}\left[H_0(r/r_*) - Y_0(r/r_*)\right], \\ r_* &= r_0/\kappa,\end{aligned} \tag{4}$$

where $C = e^2/(4\ \pi\ \epsilon_0)$, $\epsilon_0$ is the vacuum permittivity, $\kappa$ is the relative permittivity of the effective environment, and $H_0$ and $Y_0$ are the order-zero Struve and Bessel functions. The screening length $r_* = r_0/\kappa$ separates two distance regimes: the interaction varies logarithmically for $r$ much smaller than $r_*$ and approaches -$C/(\kappa\ r)$ for $r$ much larger than $r_*$ [20, 21]. It characterizes dielectric screening, independently of the strip width $L$ or the field scale introduced below.

For the offsets in Eq. (3), we adopt $\Delta\ E_\mathrm{c} = 610$ meV and $\Delta\ E_\mathrm{v} = 450$ meV from the effective band alignment reported by Huang et al. [19]. The constant scalar masses are $m_\mathrm{e} = 0.47m_0$ and $m_\mathrm{h} = 0.36m_0$, selected from the band-edge calculations of Kormanyos et al. [18]. We take $r_0 = 7.5$ nm as a phenomenological screening parameter motivated by the $WS_2$ Rydberg fit [23] and its use in interface-exciton modeling [8], and choose $\kappa = 2.5$ for the effective environment; these choices give $r_* = 3.0$ nm. Table S1 [17] specifies the provenance and screening convention. We first vary $L$, then select $L = 1.5$ nm for the eigenstate and field-response calculations in Sec. III. Parameter sensitivities are examined in Sec. S6 [17].

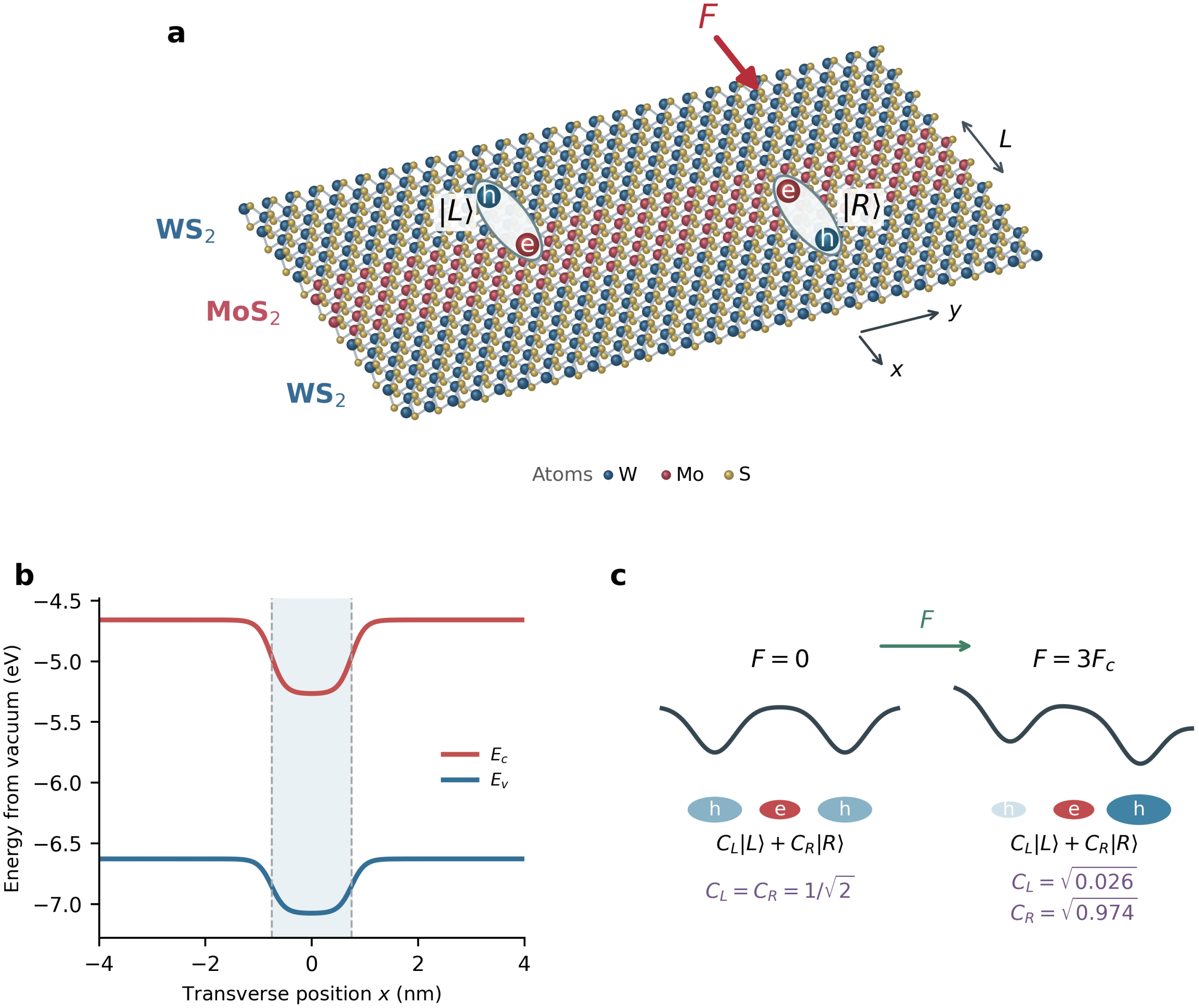

FIG. 1. One-dimensional excitons in a lateral double heterojunction. (a) Oblique view of an idealized $1H$ $WS_2$-$MoS_2$-$WS_2$ lattice with trigonal-prismatic metal coordination, elongated along $y$; the strip width $L$ is measured along $x$. W, Mo, and S sites are distinguished by color. Electron-hole pairs label the alternative one-exciton configurations $|\mathrm{L}\rangle$ and $|\mathrm{R}\rangle$ at the two interfaces; they are drawn at different longitudinal positions for clarity, not as two simultaneously present excitons. Their outlines are schematic envelopes. Atomic coordinates use a common illustrative lattice based on Ref. [18]. (b) Effective conduction- and valence-band profiles based on the UPS and optical-gap alignment in Ref. [19], with offsets of 610 and 450 meV. ($c$) Schematic CT configuration potentials and charge distributions at $F = 0$ and $F = 3F_\mathrm{c}$. The two $h$ symbols represent the spatial distribution of one hole. Here $F_\mathrm{c} = J/(ed)$, $J$ is half the zero-field even-odd splitting, $e$ is the positive elementary charge, and $ed$ is the localized-configuration dipole magnitude. Thus $edF_\mathrm{c} = J$. Both states are written as $C_\mathrm{L}|\mathrm{L}\rangle + C_\mathrm{R}|\mathrm{R}\rangle$, with real positive amplitudes for the lower branch. The displayed probabilities at $3F_\mathrm{c}$ are rounded to three decimals. These CT wells illustrate configuration energies, not the separate carrier potentials.

### B. Numerical solution

The change to center-of-mass and relative coordinates separates the kinetic energy, but the interface potentials prevent a product separation of $X$ from $(x, y)$. We therefore solve for the joint envelope by the basis expansion

$$\Psi(X, x, y) = \sum_j \chi_j(X)\phi_j(x, y).$$

Here $\phi_j(x, y)$ are eigenfunctions of the auxiliary homogeneous screened-Coulomb problem, with the interface potentials removed; $j$ labels the relative basis functions. The unknown coefficients $\chi_j(X)$ retain the dependence on transverse center-of-mass position. At each $X$, matrix elements of the band-offset potentials mix these basis functions. The $X$ kinetic term couples their coefficients at neighboring grid positions. Discretizing $X$ and diagonalizing this matrix gives the heterojunction eigenstates and eigenenergies. Both bound and positive-energy auxiliary functions are retained, so that their superposition can describe the deformed interface exciton. This is a basis expansion of the wavefunction, not a separation of $X$ or an identification of the auxiliary energies with the final exciton levels.

The basis construction, spatial grids, convergence results, and comparisons with reference systems are provided in Secs. S2-S4 of the Supplemental Material [17]. Figure S1 checks the representative doublet, Fig. S2 compares with known reference systems, and Fig. S3 details the single-interface comparison. Those comparisons distinguish numerical error for a specified Hamiltonian from differences in experimental data and in published model approximations.

### C. Binding energy

At zero field, the binding energy is the minimum energy required to separate the electron and hole infinitely far while retaining the same heterostructure potentials. The separated electron occupies the lowest transverse strip level, of energy $\epsilon_{\mathrm{e},0}$, and the hole is far away in $WS_2$, where its minimum energy is zero. Their interaction then vanishes. Therefore

$$\begin{aligned} E_{\mathrm{th}}(0) &= \epsilon_{\mathrm{e},0}, \\ E_{\mathrm{b},a} &= E_{\mathrm{th}}(0) - E_a(0). \end{aligned} \tag{5}$$

Here $E_a(0)$ is the energy of bound exciton state a and $E_{\mathrm{th}}(0)$ is the minimum energy of the infinitely separated carriers. The positive difference $E_{b,a}$ is the binding energy illustrated in Fig. 2(a). We calculate $\epsilon_{\mathrm{e},0}$ using the same electron potential and grid spacing as the two-particle calculation. In a homogeneous hydrogenic problem, this definition is equivalently $E_{n=\infty}$ - $E_{n=1}$. In the heterojunction, the corresponding limiting energy is $\epsilon_{\mathrm{e},0}$, because the separated electron still occupies the strip confinement level. The homogeneous relative-basis spectrum omits this confinement and cannot supply the binding-energy threshold.

## III. WIDTH DEPENDENCE AND EXCITON EIGENSTATES

### A. Geometric control and selection of a representative width

We first vary the $MoS_2$ strip width at fixed interface-profile width $w = 1$ nm and fixed material inputs. Figure 2(b) shows that the two lowest states remain bound throughout $L = 1.4$-$4.0$ nm. Over this interval, the even-state binding decreases from approximately 106.5 to 75.5 meV, whereas the even-odd splitting decreases from 5.605 to 0.0317 meV [Fig. 2($c$)]. The coupling $J$, defined as half this splitting, therefore changes much more strongly than the binding. Wider strips reduce the overlap of the opposite-interface configurations while preserving the attraction that binds each electron-hole pair. We define the electric crossover scale $F_{\mathrm{c}} = J/(ed)$, where $d = |\langle \mathrm{even}|x|\mathrm{odd}\rangle|$ is the dipole matrix element expressed as a length. At $F = F_{\mathrm{c}}$, the electric bias $edF$ equals the coupling $J$: the field begins to favor a single-interface configuration over an equal-weight superposition. This is a hybridization scale, not an ionization threshold. Its value decreases from 1.10 to 0.00368 V/$\mu$m [Fig. 2($d$)]. Every width uses the same production basis; numerical settings and checks are given in Secs. S3 and S6 [17].

We select $L = 1.5$ nm for the following eigenstate and electric-field calculations. $MoS_2$ strips of this width have been resolved by HAADF-STEM in lateral $MoS_2/WS_2$ structures [16], making this a structurally demonstrated geometry. The calculated gap is approximately 4.6 meV, compared with about 1.7 meV at $L = 2$ nm, while both CT1 states retain binding near 100 meV. This larger separation favors spectroscopic resolution when the linewidth is smaller than the gap. All subsequent results use $L = 1.5$ nm and $w = 1$ nm unless explicitly stated otherwise. The electric-field quantum metric provides another measure of geometric control: it quantifies how rapidly the eigenstate changes with field and gives $g_{FF} = 1/(4F_{\mathrm{c}}^2)$ at zero field in the two-state model. Its width dependence and full-wavefunction check are given in Sec. S10 and Fig. S6 [17, 24].

For very narrow strips, the two smooth interfaces overlap and reduce the actual potential contrast. Equation (2) gives $s(0) = \tanh(2L/w)$, so the central electron-well depth is $\Delta\ E_{\mathrm{c}} \tanh(2L/w)$. This reduction becomes strong for $L$ much smaller than $w$; at $L = w$ the depth is still 96.4% of its wide-strip value in our width convention. Consequently the width trend in Fig. 2 should not be extrapolated into the overlapping-interface limit. A loss of CT confinement does not necessarily remove electron-hole binding: as $L$ approaches zero, the model approaches a homogeneous $WS_2$ exciton with free transverse center-of-mass motion.

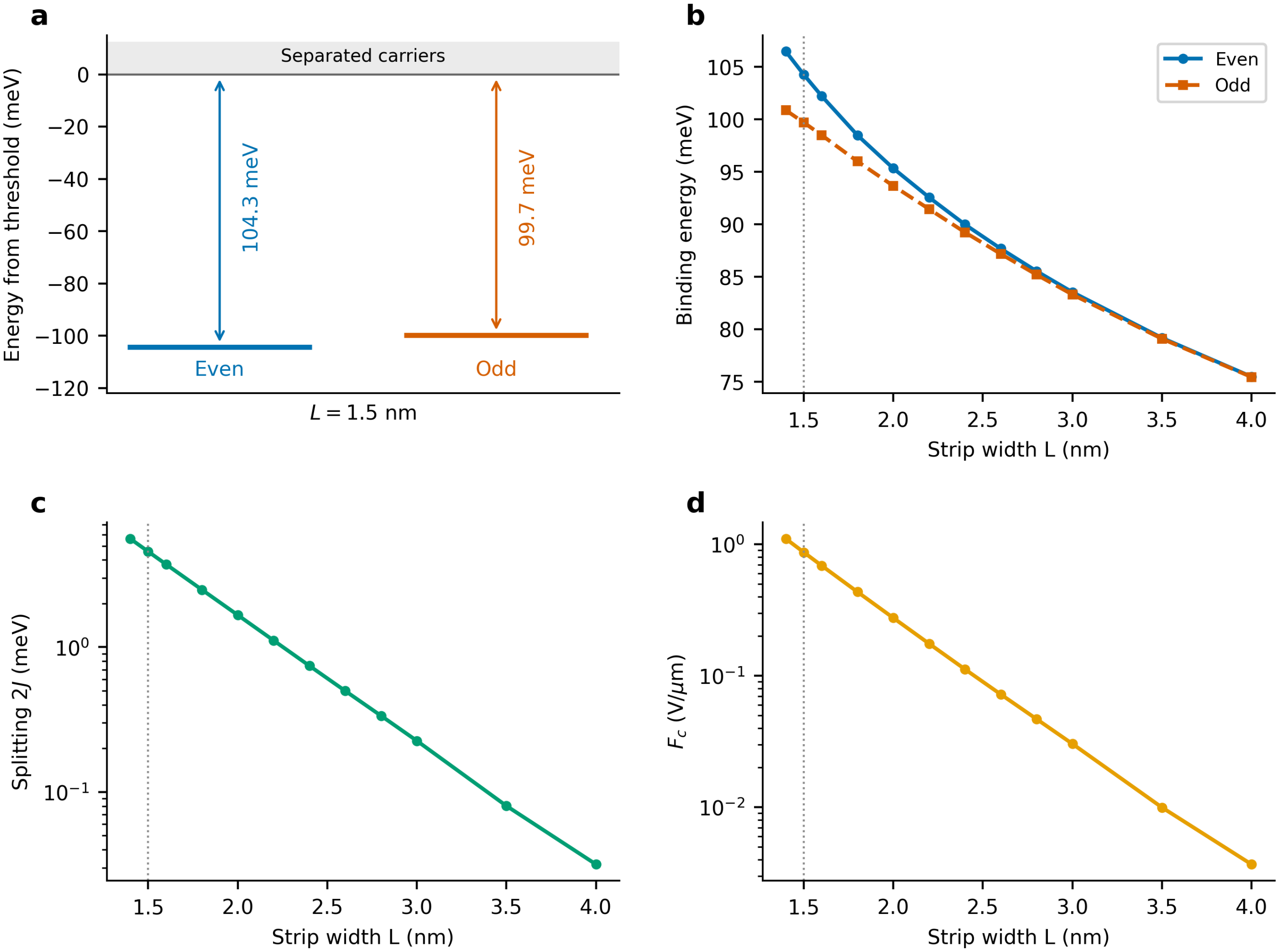

FIG. 2. Binding energies and strip-width dependence at zero field, with $w$ = 1 nm. (a) Lowest even and odd energies at $L$ = 1.5 nm relative to the separated-carrier threshold. Arrows indicate the binding energies defined by Eq. (5); this panel uses the 52-nm representation in Sec. S3 [17]. (b) Binding energies versus $L$. (*c*) Splitting $2J = E_{\mathrm{odd}}$ - $E_{\mathrm{even}}$. (*d*) Electric crossover scale $F_{\mathrm{c}} = J/(ed)$, where $d = |\langle \mathrm{even}|x|\mathrm{odd}\rangle|$. Lines in (b-*d*) connect calculations in the common 204-function basis; dotted guides mark the selected width $L$ = 1.5 nm. Numerical checks and raw data are given in Secs. S3 and S6 [17].

### B. Eigenenergies and spatial structure at $L$ = 1.5 nm

Figure 2(a) places the lowest two zero-field states below the separated-carrier energy. For $L$ = 1.5 nm and $w$ = 1 nm, the even and odd states are bound by approximately 104.3 and 99.7 meV, respectively. The odd state lies 4.6 meV above the even state; both remain well below the separated-carrier threshold. Inter-interface hybridization therefore acts on a much smaller energy scale than electron-hole binding.

Figure 3 resolves both parity and internal size. The upper row shows CT1 and the lower row the next bound pair, CT2, whose optical response is discussed in Sec. IV. Panels (a,b,$d$,$e$) show real wavefunction sections at $y = y_{\mathrm{e}}$ - $y_{\mathrm{h}} = 0$, plotted against the transverse electron and hole positions. The two colors distinguish opposite signs, or a relative phase of $\pi$. Reflection through the mirror plane $x = 0$ at the center of the $MoS_2$ strip reverses $x_{\mathrm{e}}$ and $x_{\mathrm{h}}$ simultaneously: the even envelope preserves its sign, whereas the odd envelope reverses it. CT2 also has internal nodes, distinct from this overall parity. Panels ($c$,$f$) give the radial separation probability $P_{r,a}(r)$, integrated over the center-of-mass coordinate and relative angle, including the area factor $r$. The center-of-mass probability distributions are shown in Fig. S8 [17].

We quantify the internal size by the effective exciton radius $r_{\mathrm{rms},a}$, the square root of the mean squared electron-hole separation, following the radius convention of Ref. [8]. This provides a defined size measure for an anisotropic CT exciton without fitting a hydrogenic Bohr-radius parameter. With the same masses and screened interaction but with the band-offset potentials removed, the homogeneous 2D 1$s$ reference has $r_{\mathrm{rms}}$ =

1.75 nm. The CT1 even and odd radii are 3.40 and 3.64 nm; CT2 reaches 7.32 and 7.70 nm [Fig. 3($c$,$f$)]. For localized CT1, the radius combines a mean charge separation $d = 2.63$ nm with relative-coordinate standard deviations of 1.36 nm perpendicular and 1.90 nm parallel to the interfaces (Sec. S15 [17]).

The enlarged CT1 envelope extends over distances comparable to and greater than the 1.5-nm strip width, while the electron remains shared across the central region. This spatial extent, together with finite wavefunction penetration into the intervening region, supports the coupling between opposite-interface configurations. Its strength is the Hamiltonian matrix element $J$ extracted below from the even-odd splitting, connecting the internal exciton structure to the width-controlled hybridization in Fig. 2($c$). The transverse wavefunction tails and nodal structure determine the magnitude of $J$; the total rms radius alone does not order the couplings of different CT pairs, as the larger but more weakly coupled CT2 illustrates in Sec. IV A. The internal radius exceeds the 0.549-nm interface transition distance and is determined jointly by confinement and screened attraction, not by that transition distance alone.

We characterize the material distribution by integrating the electron and hole marginal probabilities over the two constituent regions, using $s(x)$ for the smooth boundary. The probability for the electron to occupy $MoS_2$ is 0.856 (0.856) for CT1 even (odd), while the probability for the hole to occupy $WS_2$ is 0.957 (0.983). These regional occupations directly show the charge-transfer character even though the permanent dipole vanishes by left-right symmetry. Definitions, regional excess charges, and the joint CT probability are given in Sec. S2 [17].

### C. Projection onto opposite dipolar configurations

The parity here is the left-right reflection of the two-particle envelope, ($X$, $x$, $y$) to (-$X$, -$x$, $y$). Reflection symmetry labels the states but does not require their energies to be degenerate. Linear combinations of the lowest even and odd states define the configurations $|\mathrm{L}\rangle$ and $|\mathrm{R}\rangle$ localized at the left and right interfaces [Fig. 1(a)]. The physical dipole is $p_x$ = -$e$ times the expectation value of $x$. The two configurations have dipoles -$ed$ and +$ed$, respectively, where $d = |\langle\mathrm{even}|x|\mathrm{odd}\rangle|$ is obtained from the calculated relative-coordinate matrix element.

In this two-state basis, $\hat{I}$ is the identity, $\hat{\rho}_x = |\mathrm{L}\rangle\langle\mathrm{R}| + |\mathrm{R}\rangle\langle\mathrm{L}|$ exchanges the configurations, and $\hat{\rho}_z = |\mathrm{R}\rangle\langle\mathrm{R}| - |\mathrm{L}\rangle\langle\mathrm{L}|$ measures their population difference. The coupling $J$ is positive in the chosen phase convention and is extracted from the zero-field splitting as $J = [E_{\mathrm{odd}}(0) - E_{\mathrm{even}}(0)]/2$. We denote the mean configuration energy by $E_0 = (E_\mathrm{L} + E_\mathrm{R})/2$. Throughout the calculations in this article, the interfaces are equivalent: their static detuning $\Delta_0 = E_\mathrm{L} - E_\mathrm{R}$ is zero. $E_0$ is then also the mean even-odd eigenenergy. With the electric bias energy $q = edF$, projection of Eq. (1) onto this pair gives

$$\hat{H}_{\mathrm{CT}} = E_0\hat{I} - J\hat{\rho}_x - q\hat{\rho}_z. \tag{6}$$

The term $E_0\hat{I}$ fixes the energy origin, -$J$ $\hat{\rho}_x$ couples $|\mathrm{L}\rangle$ and $|\mathrm{R}\rangle$, and -$q$ $\hat{\rho}_z$ describes their electric-field bias. We label the two eigenbranches even and odd by the zero-field states to which they connect continuously. These labels specify branch origin, not a conserved parity at nonzero $F$. Their energies are

$$\begin{aligned} E_{\mathrm{even}} &= E_0 - \sqrt{J^2 + q^2}, \\ E_{\mathrm{odd}} &= E_0 + \sqrt{J^2 + q^2}. \end{aligned} \tag{7}$$

For $L$ = 1.5 nm and $w$ = 1 nm with the inputs in Table S1, the calculation gives $J$ = 2.281 meV and $d$ = 2.632 nm. Once the geometry, band offsets, masses, and screening are specified, the spatial Hamiltonian determines $J$ and $d$; neither is an independent fitting parameter. In Figs. 4(a,b,$d$), the colored solid curves are calculated after restricting the finite-field Hamiltonian to the two zero-field CT1 eigenstates, whereas the open circles are obtained by diagonalizing it in the full spatial basis of 204 relative functions at 121 transverse center-of-mass positions. The curves use $J$ and $d$ extracted from the same zero-field spatial calculation as the circles. Their difference therefore measures mixing with states outside CT1, rather than a change of numerical grid; the numerical settings are specified in the caption and Sec. S3 [17]. Figure 4($c$) shows the right-interface probabilities of the two-state model.

### D. Symmetric crossover and uncoupled control

Using $F_\mathrm{c} = J/(ed)$, we write the reduced field as $u = F/F_\mathrm{c}$. Numerically $F_\mathrm{c} = J[\mathrm{meV}]/d[\mathrm{nm}]$ in V/$\mu$m. The even-derived energy shift is

$$E_{\mathrm{even}}(F) - E_{\mathrm{even}}(0) = -J\left(\sqrt{1+u^2} - 1\right), \tag{8}$$

The odd-derived shift has the opposite sign. For branch a = even or odd, $p_{x,a} = -\partial E_a/\partial F$ is its permanent dipole, and $P_{R,a}$ is its probability in the right-interface configuration $|\mathrm{R}\rangle$. The two-state expressions are

$$\begin{aligned} p_{x,\mathrm{even}} &= ed\frac{u}{\sqrt{1+u^2}}, \\ p_{x,\mathrm{odd}} &= -p_{x,\mathrm{even}}, \\ P_{\mathrm{R,even}} &= \frac{1}{2}\left(1 + \frac{u}{\sqrt{1+u^2}}\right), \\ P_{\mathrm{R,odd}} &= 1 - P_{\mathrm{R,even}}. \end{aligned} \tag{9}$$

Equation (9) gives the blue even-derived and red odd-derived dipole curves in Fig. 4(b) and the corresponding right-interface probabilities in Fig. 4($c$). Figure 4(a) shows the even-derived branch evolving from the quadratic shift - $(edF)^2/(2J)$ to an approximately linear shift at $|edF|$ much larger than $J$. Any nonzero field breaks the left-right reflection symmetry, giving unequal interface weights. For finite $J$, both configurations remain mixed; a predominantly localized state is reached only when the bias exceeds the coupling. At $F = F_{\rm c}$, $p_{x,\rm even} = ed/\sqrt{2}$ and $P_{\rm R,even} = 0.854$; the odd-derived branch has the opposite dipole and complementary probability.

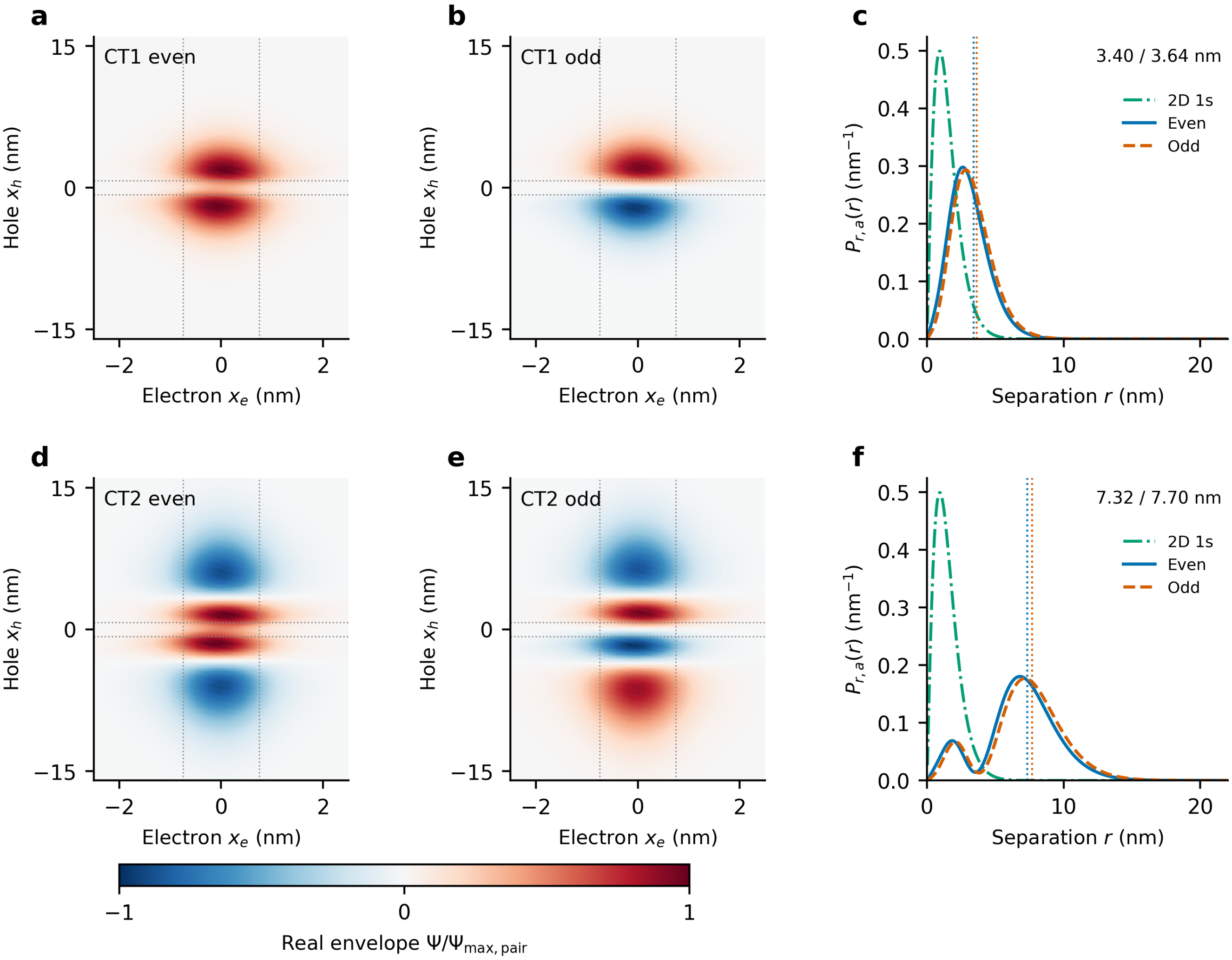

FIG. 3. Parity-resolved wavefunctions and internal sizes at $L = 1.5$ nm, $w = 1$ nm, and zero field. The upper and lower rows show CT1 and CT2. (a,$d$) Even and (b,$e$) odd real envelopes in the $y = 0$ plane, plotted against electron position $x_{\rm e}$ and hole position $x_{\rm h}$. Red and blue denote positive and negative amplitudes; a common maximum absolute amplitude rescales each parity pair. The overall sign is arbitrary. Dotted lines mark the interfaces at each carrier coordinate equal to plus or minus $L/2$. These are wavefunction sections, not marginal densities. ($c$,$f$) Radial separation distributions: even (blue solid), odd (orange dashed), and the matched homogeneous 2D 1$s$ reference (green dash-dotted). Vertical dotted lines mark the even and odd rms radii; the 2D reference radius is 1.75 nm. Radial distributions are normalized over their full domains. The 2D curve is a matched-parameter comparison, not a separate material prediction. All panels use the same 52-nm eigenstates; numerical checks and definitions are in Secs. S3, S13, and S15 [17].

The right axis of Fig. 4(b) shows the signed degree of left-right hybridization, $\langle\hat\rho_x\rangle = 2\,{\rm Re}(C_{\rm L}^* C_{\rm R})$. For the real stationary eigenstates considered here, its magnitude is one for equal-weight superpositions and zero for a single-interface state; its sign distinguishes equal and opposite relative phase. This expectation value describes the state, whereas the Hamiltonian coefficient $J$ describes the coupling that produces hybridization. The even-derived and odd-derived branches have $\langle\hat\rho_x\rangle = +J/\sqrt{J^2+q^2}$ and -$J/\sqrt{J^2+q^2}$, respectively. The left axis gives the permanent dipole in Debye. The localized-configuration magnitude $ed$ is approximately 126.4 D, about 3.8 times the 0.7 $e$ nm (34 D) scale reported for the vertical quadrupolar-exciton system in Ref. [12]. Thus the lateral geometry supports a large in-plane dipole whose sign and magni-tude can be varied electrically through the quadrupolar state.

The two-state dipoles approach plus or minus $ed$ because this projection retains the zero-field internal charge distributions: the field changes only the left-right weights. Higher-state admixture also deforms each localized configuration, allowing the full dipole to exceed this value. Thus the asymptote in Fig. 4(b) is a property of the fixed-basis model, not a maximum dipole before dissociation. Within the CT1 two-state approximation, the zero-field differential response is $\alpha_0 = \partial p_{x,\mathrm{even}}/\partial F = (ed)^2/J$, equal to the magnitude of the even-derived Stark curvature, whereas the dipole itself is zero. Higher states add approximately 1% to this response in the representative full calculation (Sec. S14 [17]). Narrowing the strip increases the splitting and therefore trades enhanced spectral separation for reduced low-field sensitivity; the width-controlled quantities satisfy $\alpha_0 F_\mathrm{c} = ed$. For $L =$ 1.5 nm, $\alpha_0$ is approximately 146 D/(V/$\mu$m). At $L =$ 2.0 nm it reaches approximately 529 D/(V/$\mu$m), about 270 times the curvature-derived response at the compensation point of the representative vertical-system fit in Ref. [12]. This comparison uses the local electric field in each model and reflects both the larger dipole length and smaller $J$ of the lateral junction (Sec. S14 [17]).

The $J = 0$ controls in all four panels isolate the effect of inter-interface coupling. In Fig. 4(a), uncoupled energies cross at zero field, whereas finite $J$ opens the avoided crossing. In Fig. 4(b), an arbitrarily small field gives the uncoupled states their full opposite dipoles and zero hybridization; finite $J$ produces a continuous dipole response. In Fig. 4($c$), the corresponding discontinuous interface selection becomes a smooth redistribution. In Fig. 4($d$), each uncoupled localized state carries half the zero-field bright-state optical strength for any nonzero field, whereas finite $J$ transfers strength continuously from the bright even-derived branch to the odd-derived branch. At the exact $J = 0$, $F = 0$ degeneracy, the individual eigenbasis is arbitrary, although the total optical strength is fixed. This control removes left-right coupling; the higher-state test below instead retains $J$ and tests the need for states beyond CT1.

### E. Field-induced mixing beyond CT1

We test whether the two zero-field CT1 eigenstates provide a sufficient basis for the electric-field response. In the two-state calculation, we diagonalize the finite-field Hamiltonian using only these even and odd states, retaining their computed coupling $J$. In the full calculation, we diagonalize the same Hamiltonian in the complete numerical spatial basis, which also represents higher CT states, Rydberg-like states, and finite-box continuum states. This comparison tests the restriction to CT1 separately for energies, permanent dipoles, and optical transition strengths. Equations (7)-(9) describe the energies, dipoles, and interface probabilities within this two-state approximation.

In the weak-field regime, at $F = F_\mathrm{c} = 0.867$ V/$\mu$m, the full calculation has weights outside the zero-field CT1 pair of 0.028% and 0.060% for the even-derived and odd-derived branches, respectively. Their additional energy lowerings are 15.3 and 30.0 $\mu$eV. The CT1 pair therefore describes the weak-field energies and leading dipole response; the direct comparisons are shown in Figs. 4(a,b) and S4 [17].

In the stronger-field regime, at $F = 3F_\mathrm{c} = 2.60$ V/$\mu$m, the corresponding weights increase to 0.40% and 0.48%, and the additional energy lowerings are 0.188 and 0.230 meV. The growing admixture also changes the internal charge distribution and hence the permanent dipole [Fig. S4(b)]. These values use the relative-coordinate domain $r \leq 36$ nm and transverse center-of-mass range $|X| \leq 20$ nm (Secs. S3 and S5 [17]).

Optical transition strengths are more sensitive than energies to higher-state mixing because their amplitudes interfere. We therefore retain the full spatial basis for the CT1 and CT2 optical resonances in Sec. IV, where isolated-pair and larger-basis calculations are compared explicitly.

## IV. EXCITED CT DOUBLETS AND OPTICAL SIGNATURES

### A. Excited bound states

Beyond CT1, the calculation resolves an excited even-odd CT pair, denoted CT2. Using a relative-coordinate domain $r \leq 52$ nm and a transverse center-of-mass range $|X| \leq 28$ nm (Sec. S3 [17]), its binding energies are 54.98 and 52.75 meV, and its splitting is 2.232 meV. The dipole matrix element $d = 5.99$ nm gives $F_{c,2} = 0.186$ V/$\mu$m, about 4.7 times smaller than the CT1 scale. Figure 5 shows its calculated field evolution and optical activation together with CT1.

The CT2 envelopes are spatially more extended, with rms electron-hole radii of approximately 7.3-7.7 nm compared with 3.4-3.6 nm for CT1 [Fig. 3($c$,$f$)]. Their radial distributions acquire an additional inner feature and a dominant outer peak. CT1 and CT2 are numbered by increasing pair energy within the sector even under relative-coordinate reflection $y$ to -$y$. Each pair contains one state even and one state odd under left-right reflection ($X$, $x$) to (-$X$, -$x$); the $y$-reflection condition is a separate symmetry label. The junction couples angular and radial components, so CT2 does not imply a pure hydrogenic 2$s$ state. This resolved excited pair extends the opposite-dipole, even-odd structure beyond CT1, providing an excited quadrupolar CT configuration. Its zero-field energies, sizes, and domain-convergence checks are given in Secs. S9, S11, and S15 [17].

Each pair $n$ has its own coupling $J_n$, given by half its zero-field even-odd splitting, and its own dipole matrix element $d_n$. Its isolated-pair response follows Eqs. (6)-

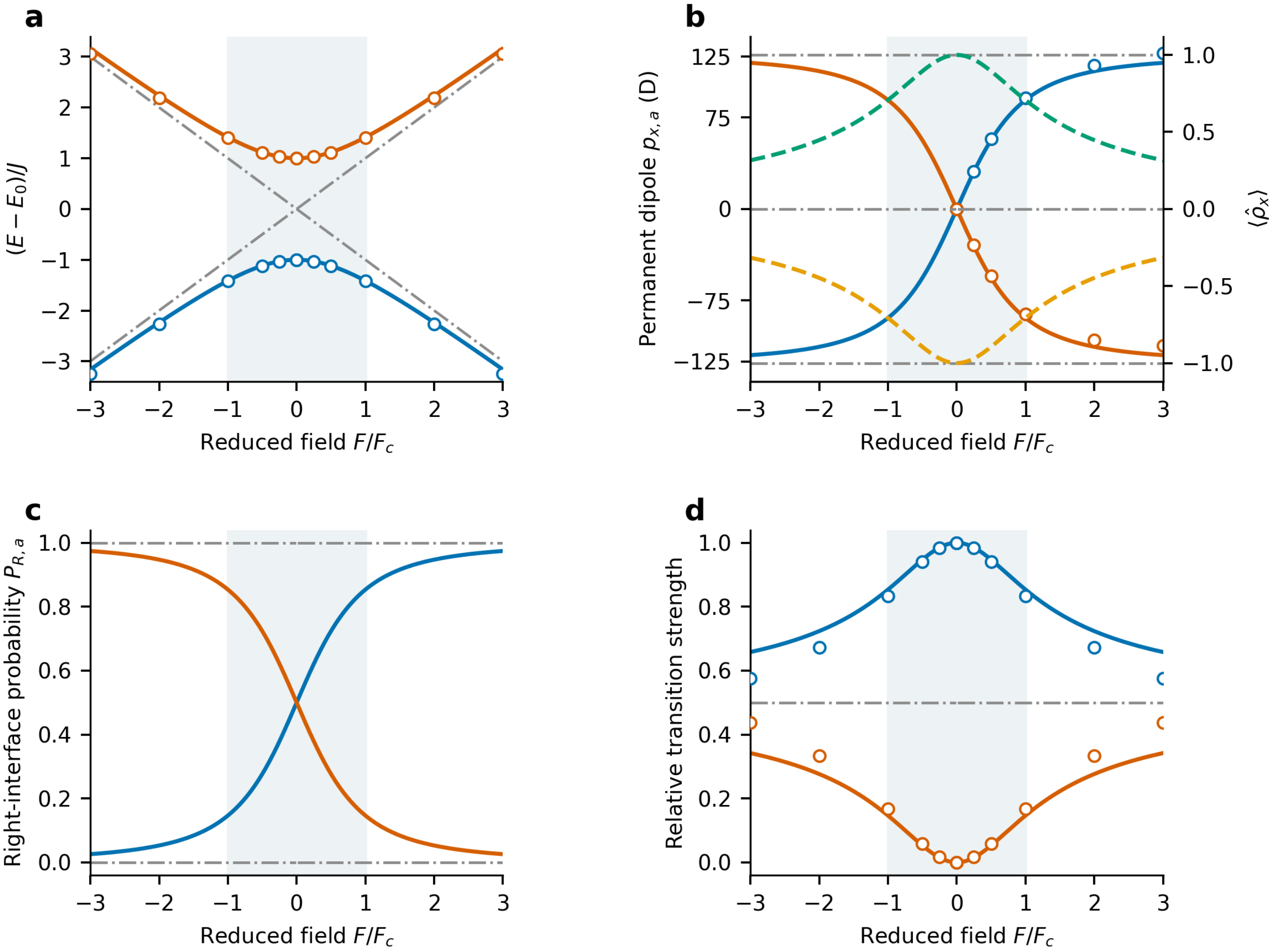

FIG. 4. Electric-field response at $L = 1.5$ nm and $w = 1$ nm. All horizontal axes are $u = F/F_c$. Blue and red solid curves denote the even-derived and odd-derived branches of the finite-$J$ two-state model, respectively; even and odd label their zero-field origin. (a) Energies relative to $E_0$, in units of $J$. (b) Permanent dipoles $p_{x,\text{even}}$ and $p_{x,\text{odd}}$ from Eq. (9) on the left axis, in Debye, and signed left-right hybridization on the right axis (green even-derived and orange odd-derived dashed curves). (*c*) Right-interface probabilities $P_{\text{R,even}}$ and $P_{\text{R,odd}}$ from Eq. (9). (*d*) Optical transition strength normalized to the zero-field bright state, assuming equal interface optical amplitudes. Open circles in (a,b,*d*) are spatial two-particle calculations, with circles in (b) showing dipoles. Negative-field points in (a,*d*) follow by reflection. $J = 2.28542$ meV, $d = 2.63000$ nm, and $F_c = 0.86898$ V/$\mu$m are extracted from the same zero-field basis as the circles. Gray dash-dotted curves are uncoupled $J = 0$ controls with fixed localized dipoles plus or minus $ed$, using the finite-$J$ reference scales; state-specific values are omitted at the exact degeneracy. They are not dissociation limits. Shading marks $|F| < F_c$.

(11) with these pair-specific parameters, so the relevant field scale is $F_{c,n} = J_n/(ed_n)$. The smaller $J$ and larger $d$ of CT2 both lower this scale. Thus a single applied field can leave CT1 strongly hybridized while driving CT2 toward a dipolar configuration. Figure S7 and Table S13 [17] compare these scales for CT1 and CT2 and their numerical-domain dependence.

A larger internal radius does not necessarily imply a larger inter-interface coupling. The enlarged CT2 radius is accompanied by outward redistribution of the hole, away from the central barrier. The average hole probability in the $MoS_2$ strip is 3.00% for CT1 and 1.34% for CT2 (Sec. S12 [17]). The lower central-barrier weight is consistent with weaker coupling despite the larger internal extent. $J$ is an off-diagonal Hamiltonian matrix element between orthonormal localized configurations; their ordinary overlap is zero and their rms radius alone does not determine $J$.

### B. Optical transition dipoles and field-induced brightening

The optical transition dipole involves coherent addition of the electron-hole coincidence amplitude over the ribbon [8]. For a state normalized over a longitudi-nal length $L_y$, and a spatially constant Bloch transition dipole $d_{\rm cv}$ projected onto the incident polarization, the transition dipole $D_a$ from the no-exciton ground state to exciton state a and its envelope factor $S_a$ are

$$
\begin{aligned}
D_a &= d_{\rm cv}\sqrt{L_y}\int dX\,\Psi_a(X,0),\\
S_a &= \left|\int dX\,\Psi_a(X,0)\right|^2,
\end{aligned}
\tag{10}
$$

The quantity $|D_a|^2$ is the dipole factor entering Fermi's golden rule for absorption. We plot the relative transition strength $S_a/S_{\rm CT1,even}(0)$; the common Bloch dipole and longitudinal normalization cancel. Absolute transition dipoles can be obtained once $d_{\rm cv}$ and the longitudinal normalization are specified. At zero field, the even envelope adds across the interfaces, while the odd envelope cancels. The finite local probability of electron-hole coincidence does not prevent this optical cancellation; its distinction from the transition dipole is detailed in Sec. S7 [17].

Using $q = edF$ and normalizing to the zero-field bright-state strength, the ideal relative transition strengths $f_{\rm even}$ and $f_{\rm odd}$ of the two branches are

$$
\begin{aligned}
f_{\rm even} &= \frac{1}{2}\left(1+\frac{J}{\sqrt{J^2+q^2}}\right),\\
f_{\rm odd} &= \frac{1}{2}\left(1-\frac{J}{\sqrt{J^2+q^2}}\right).
\end{aligned}
\tag{11}
$$

Figure 4($d$) compares this transfer with the full calculation. At $F_{\rm c}$, the odd-derived relative strength is 0.146 in the two-state expression and approximately 0.168 in the spatial calculation. The field mixes the zero-field even and odd states, removes the exact cancellation of the odd envelope, and makes the odd-derived branch optically active. In the ideal two-state limit at $|edF|$ much larger than $J$, both eigenstates approach single-interface configurations and each carries half the zero-field bright-state strength. This is an optical consequence of localization; parity itself is no longer a good quantum number at finite field.

Figure 5(a) shows how field-induced brightening and avoided crossings recur in the excited spectrum, while Figs. 5(b,$c$) resolve the corresponding CT1 and CT2 spectral features. These linked energy and intensity changes provide signatures that may be observed in reflectance spectroscopy. Quantitative reflectance contrast follows by embedding the calculated exciton susceptibility in the optical response of the substrate stack; the displayed maps show the underlying discrete exciton resonances.

### C. Optical redistribution beyond isolated pairs

At $F = 0.85$ V/$\mu$m, the full CT2 even-derived and odd-derived relative strengths are 0.232 and 0.254, whereas the isolated CT2 projection gives 0.281 and 0.182. Retaining CT1 and CT2 together gives 0.289 and 0.174; it does not recover the full redistribution. These values use the 44-nm representation, with the 36-nm comparison in Sec. S11 and Fig. S7 [17]. Thus the excited optical spectrum calls for higher states even when the lowest doublet is adequate for the leading switching response.

The optical response is especially sensitive to this admixture because the transition amplitudes add coherently. A small additional wavefunction component changes the intensity through interference with the original amplitude, to first order in its amplitude, whereas its probability weight is second order. Consequently, an accurate two-state description of the CT1 energies does not imply that independent doublets reproduce the excited-state optical spectrum. The field window in Fig. 5 already exposes this distinction through intensity transfer involving higher exciton states, while the CT2-derived states remain spatially localized in the tested domains.

## V. DISCUSSION AND CONCLUSION

The lateral geometry gives a means of varying hybridization through an in-plane length while retaining a bound CT state. This distinguishes the material implementation and geometric control from the established vertical quadrupolar-exciton mechanism [12, 15]. The observation of a quadratic Stark shift alone would not uniquely identify such a state: other polarizable bound excitons can also shift quadratically. The stronger test is the combination of an avoided crossing, a field-induced dipole approaching the single-interface value, redistribution of optical transition strength, and a systematic dependence of the gap on interface separation.

The effective-mass description separates geometric control from microscopic material inputs. Band offsets, effective masses, and screening determine the quantitative scales, while the strip width controls the spatial overlap of the two interface configurations. The excited bound pairs show that this organization extends beyond CT1. The lowest pair captures the weak-field energy and dipole response, while higher-state mixing produces observable redistribution of the excited-state optical transition strengths. Parameter sensitivities are summarized in Sec. S6 and Fig. S5 [17].

A static left-right energy difference $\Delta_0$ need not remove this mechanism. Within the two-state model with $J$ and $d$ held fixed, replacing $q$ by $edF + \Delta_0/2$ shifts the compensation field to $F_{\rm sym}$ = -$\Delta_0/(2ed)$, while preserving the minimum gap $2J$ and the conversion between hybridized and interface-polarized states. Microscopically inequivalent interfaces can also modify $J$ and the dipole and optical matrix elements [25, 26]; the detuning-only result isolates the shift of the control field (Sec. S7 [17]).

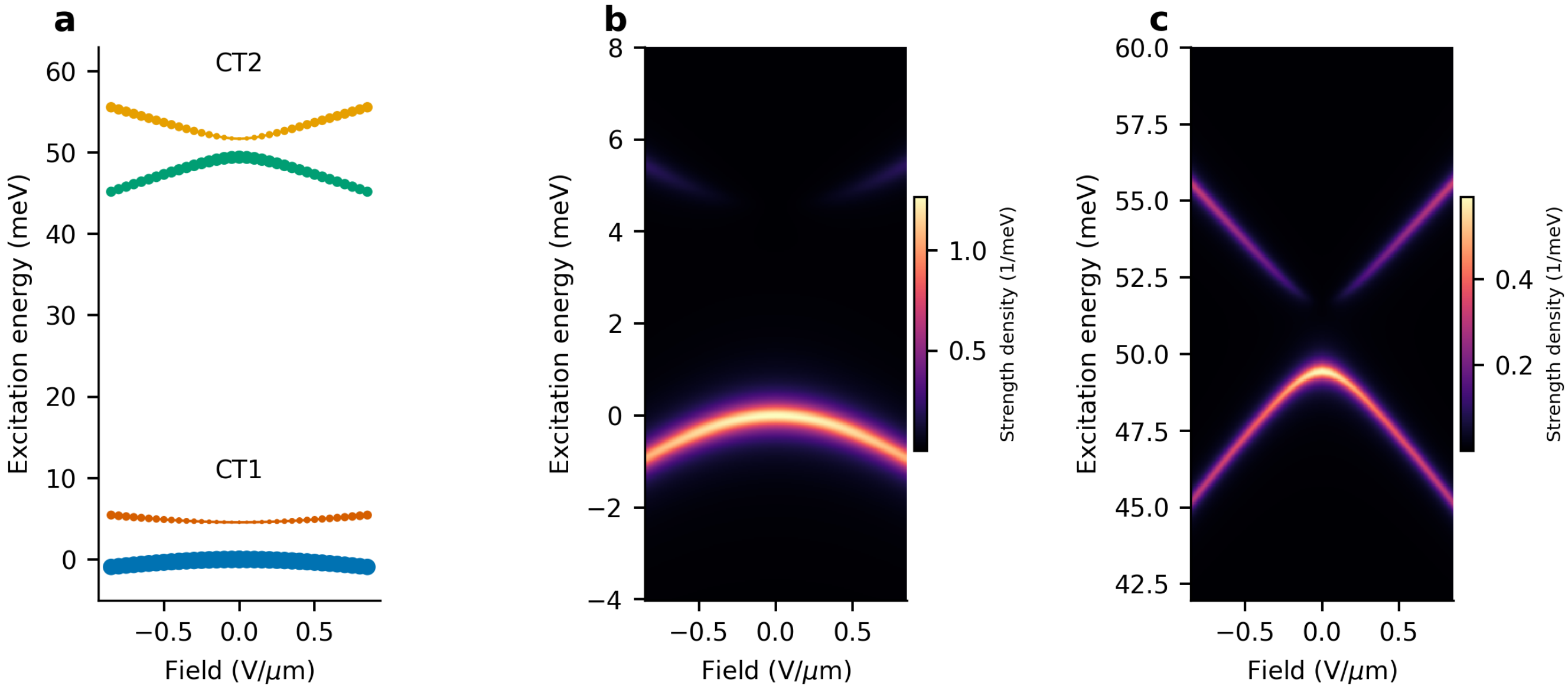

FIG. 5. Excited CT states and optical response for $L = 1.5$ nm and $w = 1$ nm. (a) CT1 and CT2 levels versus transverse electric field, with marker area proportional to the relative optical transition strength. Negative-field points are obtained by reflection. (b,$c$) Optical-strength-weighted spectra around CT1 and CT2, respectively. A Lorentzian half-width $\gamma = 0.25$ meV is used for plotting. Energies are measured from the zero-field CT1 even state; an absolute interband gap is not added. All panels use full spatial eigenstates in the enlarged numerical domain described in Sec. S9 [17], rather than assigning excited levels from a two-state Hamiltonian. The maps contain these four levels, with energies and weights interpolated between the sampled fields; no continuum background is included.

In summary, our spatially resolved two-particle calculation establishes bound one-dimensional quadrupolar excitons in a lateral double heterojunction. At zero field, coherent coupling of opposite-interface CT configurations produces split parity states with zero permanent dipole and nonzero charge quadrupoles. The representative $L = 1.5$ nm strip combines binding near 100 meV and a 4.6-meV splitting with a large low-field Stark curvature: the CT1-projected differential dipole response is approximately 146 D/(V/$\mu$m), about 75 times that of the representative vertical quadrupolar-exciton fit [12]. At $L =$ 2.0 nm, this ratio reaches approximately 270, illustrating control through strip width. A transverse field induces a large in-plane dipole on the 126-D localized-configuration scale for $L = 1.5$ nm, while continuously redistributing the optical transition strength. The calculated avoided crossing and field-induced brightening of the odd-derived branch, together with the excited CT2 resonances in Fig. 5, provide spectroscopic signatures for experimental detection. Nanometer-wide lateral strips have already been fabricated [16]; resolving the predicted doublet requires an optical linewidth below its meV-scale splitting. These results identify lateral double junctions as a platform for geometrically tunable and electrically switchable one-dimensional quadrupolar excitons.

### A. Future prospect

The states $|\mathrm{L}\rangle$ and $|\mathrm{R}\rangle$ also define a charge-configuration pseudospin, analogous to the localized basis used for double-quantum-dot charge qubits [27]. This is a spatial degree of freedom, distinct from the spin and valley labels not resolved by the scalar envelope Hamiltonian. In Eq. (6), strip geometry sets the coupling $J$, while an applied field controls the bias $q$; pulsed bias control therefore offers a route to rotations about distinct pseudospin axes at a designed $J$. Exploring this degree of freedom for quantum information will require single-exciton preparation, longitudinal confinement for individually addressable sites, coherent manipulation and readout within the exciton lifetime, and controlled inter-exciton interactions for two-qubit operations.

## ACKNOWLEDGMENTS

R. K. acknowledges support from JSPS KAKENHI under Grant Nos. JP25K22210, JP25K24588, JP24$H$02218, and JP23$H$05469, and from the World Premier International Research Center Initiative (WPI), MEXT, Japan.

OpenAI Codex was used to assist with coding and testing, figure preparation, and manuscript editing. All AI-assisted text was reviewed and revised by the author, who takes full responsibility for the final manuscript, including its scientific interpretation and verification.

## DATA AVAILABILITY

The numerical data underlying the figures, calculation and plotting scripts, and model parameters are included in the accompanying data and code archive described in Sec. S8 of the Supplemental Material [17].

# Supplemental Material

Electrically switchable one-dimensional quadrupolar excitons in lateral double heterojunctions

Ryo Kitaura

Research Center for Materials Nanoarchitectonics, National Institute for Materials Science, 1-1 Namiki, Tsukuba, Ibaraki 305-0044, Japan

Graduate School of Chemical Sciences and Engineering, Hokkaido University, 5, Kita 8 Nishi, Kita-ku, Sapporo, Hokkaido 060-8628, Japan

KITAURA.Ryo@nims.go.jp

Reference numbers refer to the bibliography of the main article.

## S1. Model inputs and their provenance

Table S1 distinguishes literature-derived inputs from parameters selected for the present model. $J$, $d$, the binding energy, and the crossover field are outputs rather than independently chosen material inputs.

Table S1. Model inputs and provenance. Reference numbering follows the main article. The screening convention is $r_* = r_0/\kappa$.

| **Parameter** | **Value** | **Source and use in this model** |
|---|---|---|
| $m_e/m_0$ | 0.47 | Ref. [18], Table 3: $MoS_2$ K conduction branch (1), PBE/PBE. Adopted as a constant scalar electron mass. |
| $m_h/m_0$ | 0.36 | Ref. [18], Table 4: magnitude of the $WS_2$ upper K-valence-band mass, PBE/PBE. Adopted as a constant scalar hole mass. |
| $\Delta\ E_c$ | 610 meV | Ref. [19], band-alignment discussion associated with Fig. 5 and Fig. S17: conduction offset inferred from UPS valence alignment and PL gaps. |
| $\Delta\ E_v$ | 450 meV | Ref. [19], same alignment: the $MoS_2$ valence edge is lower than the $WS_2$ edge by 0.45 eV. |
| $L$ | 1.5 nm | Representative width selected in Fig. 2; HAADF-STEM demonstration in Ref. [16], Fig. 5(a). |
| $w$ | 1.0 nm | Smooth-profile choice in Eq. (2), not a measured band-edge transition width; 10%-90% distance 0.549 nm. Varied in Fig. S5. |
| $\kappa$ | 2.5 | Effective-environment parameter; varied from 2.0 to 3.0 in Fig. S5. |
| $r_0$ | 7.5 nm | Phenomenological reference value inherited from the 75-Angstrom fit in Ref. [23] and its use in Ref. [8]; the convention issue is explained below. |

The electron mass corresponds specifically to branch (1) of Ref. [18], not to a spin-resolved assignment of the lowest optically allowed CT transition. This calculation treats the masses as scalar envelope parameters; their sensitivity is tested without assigning a Bloch spinor to the optical amplitude.

The screening parameter requires particular care. Ref. [23] fitted a 75-Angstrom length in its $WS_2$ Rydberg model and stated that it partly incorporates substrate screening. It is therefore not a measured intrinsic polarizability length of the present heterojunction. We retain 7.5 nm as a phenomenological reference in Eq. (4), with $\kappa$ introduced separately and $r_* = 3.0$ nm for the selected inputs. This convention specifies the model interaction; it is not a unique conversion of the experimental fit into an intrinsic material constant. The independent $r_0$ and $\kappa$ sweeps in Fig. S5 test its influence. The short-distance limit is $V(r)$ approximately $C[\ln(r/(2r_*)) + \gamma_E]/(\kappa\ r_*)$, with Euler's constant $\gamma_E$; at large distance $V(r)$ approaches -$C/(\kappa\ r)$. Thus $r_*$ separates logarithmic from Coulomb-like screening.

Figure 1(b) retains the effective band edges used in the alignment of Ref. [19]: $E_\mathrm{c}$ and $E_\mathrm{v}$ are -5.27 and -7.08 eV for $MoS_2$ and -4.66 and -6.63 eV for $WS_2$, relative to vacuum. The conduction edges involve optical gaps, so these are effective alignment inputs rather than quasiparticle energies. No additional contact-induced electrostatic profile is included. The two offsets, not the common absolute energy reference, enter Eq. (3).

For plotting, Fig. 5(b,c) uses a Lorentzian half-width $\gamma = 0.25$ meV. Figure 4 uses the common 204-function normalization of Sec. S3. Figure 3 uses the 52-nm representation, and Fig. 5 the 36-nm representation. All representative calculations in this supplement use $L = 1.5$ nm unless explicitly specified otherwise.

The structural length scales in Ref. [16] must be distinguished. Its HAADF-STEM images show atomically sharp chemical interfaces and $MoS_2$ strip widths of 3, 1.5, and 0.8 nm, as well as a single atomic row. These are strip widths, corresponding to $L$, rather than interface transition widths. Figure S8 of that reference reports a 1.6-nm penetration length from an exponential fit associated with the CBM wavefunction tail. This is a wavefunction property, not a fit to a conduction- or valence-band edge. The paper therefore motivates sharp-interface and nanometer-strip modeling but does not supply a numerical 10%-90% electronic transition length to substitute for $w$. The choice $w$ = 1 nm is independent of $L$; a band-edge profile extracted from spatially resolved electronic-structure data would determine that separate input.

The atomic illustration in Fig. 1(a) uses the $1H$ trigonal-prismatic structure described in Ref. [18], with a common in-plane lattice constant 0.318 nm and S-plane separation 0.314 nm, rounded from the PBE structural values in its Table 1. These coordinates are used only for drawing. They do not replace the smooth continuum potentials or imply reflection-equivalent relaxed atomic interfaces.

## S2. Discretization and observable definitions

### A. Longitudinal wave vector and retained coordinates

For constant electron and hole masses, the transformation to center-of-mass coordinates $(X,Y)$ and relative coordinates $(x,y)$ separates the kinetic energy exactly. The potentials depend on $X$, $x$, and $y$ but not on $Y$. Consequently the operator -i $\hbar\ \partial_Y$ commutes with the Hamiltonian; its eigenvalue is $\hbar\ K_y$, the total envelope momentum parallel to the interfaces. The total wavefunction and dispersion can be written as

$$\Phi_{n,K_y}(X,Y,x,y) = \frac{e^{iK_yY}}{\sqrt{L_y}}\Psi_n(X,x,y), \qquad E_n(K_y) = E_n(0) + \frac{\hbar^2 K_y^2}{2M}. \tag{S1}$$

Here $n$ labels an internal exciton state. The longitudinal plane wave is normalized over length $L_y$, consistently with Eq. (10). We calculate $K_y = 0$. The relative coordinate $y = y_\mathrm{e}$ - $y_\mathrm{h}$ remains in $\Psi_n$; neither carrier is constrained to have zero individual longitudinal momentum. No product separation of $X$ and $r$ is made in the spatial two-particle solution.

### B. Relative motion basis and the radial boundary

We first remove the interface potentials and solve the auxiliary homogeneous problem $\hat{h}_\mathrm{rel}$ = -$\hbar^2\nabla_r^2/(2\mathrm{mu}) + V(r)$. In polar relative coordinates, $x = r\ \cos(\theta)$ and $y = r\ \sin(\theta)$, the basis functions are $\phi_{nm}(r,\theta) = R_{nm}(r)A_m(\theta)$. The radial functions satisfy

$$\left[-\frac{\hbar^2}{2\mu}\left(\frac{d^2}{dr^2} + \frac{1}{r}\frac{d}{dr} - \frac{m^2}{r^2}\right) + V(r)\right] R_{nm}(r) = \epsilon_{nm} R_{nm}(r). \tag{S2}$$

The index $n$ enumerates radial eigenfunctions for a given nonnegative integer angular index $m$. We use $A_0 = 1/\sqrt{2\pi}$ and $A_m = \cos(m\ \theta)/\sqrt{\pi}$ for $m \geq 1$. Thus an angular cutoff $m_{max} = 10$ means all eleven channels $m$ = 0,1,...,10, not just cos(10 $\theta$). These channels span states even under relative $y$ reflection. The sine channels decouple from them under the model potentials and a transverse field. Odd-in-$y$ states have zero electron-hole coincidence amplitude and do not contribute to normal-incidence dipole absorption in the present local optical model. The reported excited spectrum is the even-in-$y$ sector, not an enumeration of every dark exciton.

The radial box is the finite numerical range $0 \leq r < R_{max}$ for the electron-hole separation, not the physical ribbon width. The production representation uses $R_{max}$ = 28 nm and $N_r$ = 560 cell centers, $r_i$ = (i + 1/2) $\Delta$ $r$ with $\Delta$ $r$ = 0.05 nm and i = 0,...,559. The radial finite-volume representation uses $\sqrt{r_i \Delta r}$ R($r_i$) to represent the measure $r$ dr. Regularity is imposed at the origin and a zero exterior value terminates the outer mesh. Increasing $R_{max}$ checks that this artificial boundary does not control a localized state.

We generate up to 20 radial functions for each angular channel and retain those with auxiliary energy $\epsilon_{nm} \leq 1000$ meV. Negative $\epsilon_{nm}$ describes an auxiliary bound state; the finite radial range also discretizes the positive-energy sector into normalizable pseudostates. These are basis functions that help represent a deformed bound CT exciton, not evidence that it has ionized. The production representation retains 204 relative basis functions after the energy cutoff.

## $C$. Transverse grid and coupled matrix problem

Writing $j$ for the combined radial and angular basis index, the remaining envelope is

$$\Psi(X, r, \theta) = \sum_j \chi_j(X)\phi_j(r, \theta). \tag{S3}$$

The unknown functions $\chi_j(X)$ are expansion coefficients that vary across the ribbon. Substitution in Eq. (1) gives coupled equations for these functions: the relative energies $\epsilon_j$ are diagonal, while matrix elements of $V_{\mathrm{e}}(X$ + $m_{\mathrm{h}}$x$/M)$, $V_{\mathrm{h}}(X$ - $m_{\mathrm{e}}$x$/M)$, and $eFx$ mix the relative functions. Retaining this coupling distinguishes the calculation from a single adiabatic relative state at each $X$.

The production representation $X$ grid contains $N_X$ = 121 positions, $X_i$ = -16 nm + i $h$ for i = 0,...,120, where $h$ = 32/120 nm = 0.266667 nm. This is a grid for the transverse center of mass, not a set of longitudinal wave vectors. Its kinetic operator uses the centered second difference [$\chi(X$+$h)$ - 2 $\chi(X)$ + $\chi(X$-$h)]/h^2$, with zero exterior values. Combining 121 positions with 204 relative functions gives a matrix dimension of 24684.

Interface matrix elements use 180 radial Gauss-Legendre points and 80 uniformly spaced angles in the production calculation. Sparse Hermitian diagonalization selects the lowest states with an eigensolver tolerance of $10^{-9}$. This controls algebraic solution error and does not replace checks of the basis cutoff, grid spacing, or boundary location. The separated-carrier threshold is calculated from the single-electron strip Hamiltonian on the same $X$ grid; the asymptotic hole potential is zero.

## D. Observables and interface profile width

For $X$ spacing $h$ and a normalized discrete eigenvector $c_{ia}$, the projected relative-coordinate matrix element between states $u$ and v is the sum over $X$ blocks of $c_u^T\ x\ c_{\mathrm{v}}$. No additional $h$ is needed because the discrete coefficients include $\sqrt{h}$. Reflection acts by reversing the $X$ grid and multiplying the relative angular channel by $(-1)^m$. The physical dipole is -$e$ times the expectation of $x = x_{\mathrm{e}}$ - $x_{\mathrm{h}}$. The plotted right-interface occupation is a probability in the two-state configuration basis, not an independently integrated sharp spatial partition.

For the optical quantities, only $m = 0$ contributes at $r = 0$. The implementation estimates its origin value from the first radial sample. The integrated local electron-hole coincidence density $C_a$ is the sum of squared coincidence amplitudes over the $X$ grid. The optical envelope amplitude is their sum multiplied by $\sqrt{h}$; its square gives $S_a$ in Eq. (10). Their units are $nm^{-2}$ and $nm^{-1}$, respectively. Section S7 relates these quantities to the transition dipole without identifying them with each other.

The smooth CT indicator is $s(x_e)[1 - s(x_h)]$. It distinguishes electron weight inside $MoS_2$ and hole weight outside it without imposing a sharp interface. For an isolated step $[1 + \tanh(4x/w)]/2$, the distance between 10% and 90% is $w$ atanh(0.8)/2 = $0.549306w$. Thus $w = 1$ nm is not a 1-nm 10%-90% width. Ref. [8] uses $\tanh(x/w)$, whereas Ref. [9] uses $\tanh(4x/w)$; their parameters called $w$ differ by a factor of four for the same spatial profile.

The joint CT probabilities at $L = 1.5$ nm are approximately 0.819 and 0.841 for CT1 even and odd. Reflection imposes equal left-right probabilities within each state, not equal joint CT probabilities between different states. In the localized basis, the reflection-even operator Q has equal diagonal entries and can have a nonzero off-diagonal entry. Its even and odd expectations therefore differ by an interference contribution. This is distinct from the Hamiltonian coupling $J$.

The central contrast of the double-step profile is $s(0) = \tanh(2L/w)$. Thus the electron well and hole barrier become shallow when $L/w$ is much less than one. Disappearance of a separated-interface CT configuration is not equivalent to disappearance of a Coulomb-bound exciton. At $L = 0$, the potentials become spatial constants and a homogeneous exciton survives, while its transverse center of mass is no longer confined by a strip.

### $E$. Regional carrier occupations and excess charge

The electron probability in $MoS_2$ is the expectation value of $s(x_e)$, denoted $P_{e,MoS_2}$; the hole probability is the expectation value of $s(x_h)$. Both expectations integrate the joint probability density over $X$, $x$, and $y$. Their $WS_2$ probabilities are one minus these values. The excess envelope charge in $MoS_2$ is $e(P_{h,MoS_2} - P_{e,MoS_2})$, and that in $WS_2$ is its negative. This integrates the charge of the exciton over material regions, not the ground-state all-electron density used in a Bader analysis. The smooth product Q in Sec. S2 D is a joint probability; it is distinct from either marginal probability.

Table S2. Regional probabilities and excess charges at $L = 1.5$ nm in the 52-nm representation. Values are normalized by the independently integrated quadrature norm, whose departure from one is recorded in the data. Charges are in units of $e$.

| **State** | $P_{e,MoS_2}$ | $P_{h,WS_2}$ | $MoS_2$ **charge** | $WS_2$ **charge** |
|---|---|---|---|---|
| CT1 even | 0.85555 | 0.95708 | -0.81263 | 0.81263 |
| CT1 odd | 0.85580 | 0.98294 | -0.83874 | 0.83874 |
| CT2 even | 0.85837 | 0.98020 | -0.83856 | 0.83856 |
| CT2 odd | 0.85833 | 0.99291 | -0.85124 | 0.85124 |

## S3. Convergence of the representative doublet

We retain positive-energy auxiliary relative states up to 1000 meV in every representation. The relative radial spacing is 0.05 nm and the transverse spacing is 0.266667 nm in the domain comparison. Increasing the box and angular/radial basis together checks their combined truncation error. A separate transverse-grid test at $L = 1.5$ nm is given below.

Table S3. Numerical representations and CT1 results at $L = 1.5$ nm. R and $X_{max}$ are in nm; binding and splitting are in meV.

| **R** | $m_{max}$ | $N_{basis}$ | $N_X$ | $E_{b,even}$ | $\mathbf{2}J$ |
|---|---|---|---|---|---|
| 28 | 10 | 204 | 121 | 104.27991 | 4.570843 |
| 36 | 14 | 351 | 151 | 104.29406 | 4.562744 |
| 44 | 18 | 538 | 181 | 104.29912 | 4.562850 |
| 52 | 22 | 766 | 211 | 104.30454 | 4.562656 |

The corresponding transverse half-ranges are 16, 20, 24, and 28 nm. Radial/angular quadratures are 180/80, 240/112, 300/144, and 360/176 points. From the 44- to 52-nm representation the CT1 even energy changes by 0.00542 meV and the splitting by 0.000194 meV. Figure 4 consistently uses the 28-nm parameters $J$ = 2.285421 meV, $d$ = 2.630001 nm, and $F_c$ = 0.868981 V/$\mu$m for both lines and numerical points.

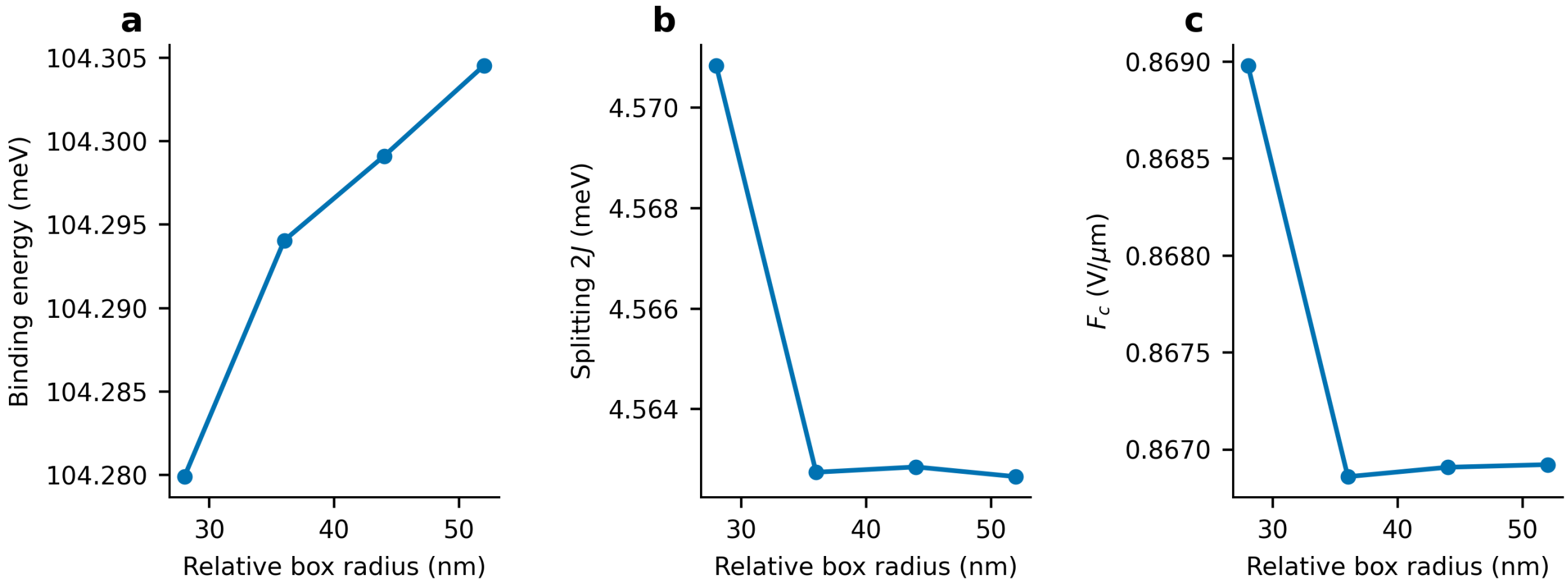

Figure S1: Representation dependence at $L = 1.5$ nm. (a) Even-state binding. (b) Even-odd splitting. ($c$) Crossover field. Basis size and transverse domain increase together as specified in Table S3; this plot is not a transverse-grid extrapolation.

Table S4. Independent transverse-grid refinement with 204 relative functions, R = 28 nm, and $X_{max}$ = 16 nm. The electron threshold is recalculated on each grid.

| $N_X$ | $h$ (nm) | $E_{b,even}$ (meV) | 2$J$ (meV) | $F_c$ (V/$\mu$m) |
|---|---|---|---|---|
| 121 | 0.266667 | 104.27991 | 4.57084 | 0.86898 |
| 181 | 0.177778 | 104.56803 | 4.62904 | 0.87662 |
| 241 | 0.133333 | 104.66953 | 4.64939 | 0.87931 |

Halving the transverse spacing changes the production-basis binding by 0.390 meV (0.37%), the splitting by 0.0785 meV (1.72%), and the crossover field by 1.19%. Thus the small changes with box size should not be mistaken for the total discretization error. Main-text scales are quoted approximately, while figure captions and CSV files identify the actual discretization-specific normalization; no fitted extrapolation is substituted for a computed eigenstate.

Additional optical checks retain 204 relative functions and compare the four CT1/CT2 states at $F$ = 0.85 V/$\mu$m, using 360 radial and 160 angular quadrature points. Halving $h$ from 0.266667 to 0.133333 nm changes their relative transition strengths by at most 0.00279. Independently halving $\Delta r$ from 0.05 to 0.025 nm at the original $h$ changes them by at most 0.000139 and changes the CT1 even binding by 0.00491 meV. At the original radial and transverse grids, doubling the production quadrature from 180/80 to 360/160 changes the four energies by less than 0.001 meV and the relative strengths by less than 0.000027. These are separate one-at-a-time checks, not a combined infinite-basis extrapolation. The numerical values are archived in audit/mesh_optical_checks.csv and audit/numerical_audit.json.

## S4. Checks against known systems

### A. Pure Coulomb problem with an exact spectrum

The 0.044% accuracy quoted for Fig. S2(a) concerns the pure interaction $V(r)$ = -$C/(\kappa\, r)$, not the Rytova-Keldysh interaction. We set $\mu = 0.20 m_0$ and $\kappa = 4.5$. Its exact negative eigenenergies are

$$E_{n_r,m} = -\frac{\mathrm{Ry}^*}{(n_r + |m| + \frac{1}{2})^2}. \tag{S4}$$

Here $n_r$ = 0,1,... is the radial index and $m$ is the angular index; the states $(n_r,|m|)$ = (0,0), (1,0), and (0,1) are 1$s$, 2$s$, and 2$p$. Ry$*$ = 1000(13.605693122994)$(\mu/m_0)/\kappa^2$ meV. Table S5 reports the energies, absolute errors $|E_{num}$ - $E_{exact}|$, and relative errors $100|E_{num}$ - $E_{exact}|/|E_{exact}|$. All cases use $R_{max}$ = 100 nm. Halving the radial spacing reduces the errors approximately fourfold, as expected for the discretization. At $\Delta$ $r$ = 0.025 nm, used in Fig. S2(a), the largest error is approximately 0.044% and the spurious 2$s$-2$p$ splitting is about 0.0039 meV. The exact spectrum has zero 2$s$-2$p$ splitting.

Table S5. Radial-grid convergence toward the exact two-dimensional Coulomb spectrum. Energies and absolute errors are in meV; $\Delta$ $r$ is in nm.

| **State** | $\Delta$ $r$ | $E_{num}$ | $E_{exact}$ | **Absolute error** | **Relative error (%)** |
|---|---|---|---|---|---|
| 1$s$ | 0.05 | -536.564303 | -537.508864 | 0.944561 | 0.175729 |
| 2$s$ | 0.05 | -59.711509 | -59.723207 | 0.011698 | 0.019586 |
| 2$p$ | 0.05 | -59.727112 | -59.723207 | 0.003905 | 0.006538 |
| 1$s$ | 0.025 | -537.272101 | -537.508864 | 0.236763 | 0.044048 |
| 2$s$ | 0.025 | -59.720282 | -59.723207 | 0.002925 | 0.004898 |
| 2$p$ | 0.025 | -59.724183 | -59.723207 | 0.000976 | 0.001633 |
| 1$s$ | 0.0125 | -537.449635 | -537.508864 | 0.059229 | 0.011019 |
| 2$s$ | 0.0125 | -59.722476 | -59.723207 | 0.000731 | 0.001224 |
| 2$p$ | 0.0125 | -59.723451 | -59.723207 | 0.000244 | 0.000408 |

## B. Screened $WS_2$ eigenstates and independent numerical check

For Fig. S2(b), we use Eq. (4) with $\mu = 0.16m_0$, $\kappa = 1$, and $r_0 = r_* = 7.5$ nm. This reproduces the fitted potential convention of Chernikov et al. [23]; its length already partly accounts for substrate screening. The optical-reference gap $E_g$ = 2.41 eV converts negative relative eigenenergies $\epsilon_{ns}$ into transition energies $E_g + \epsilon_{ns}/1000$, when $\epsilon$ is in meV. No interface potentials are included. Table S6 gives all five calculated $s$ states and the rounded spectral readings used in Fig. S2(b). These readings are experimental peak estimates, not tabulated exact eigenvalues of the fitted model.

Table S6. Rytova-Keldysh $WS_2$ series at $R_{max}$ = 200 nm and $\Delta$ $r$ = 0.025 nm. The final column is the calculated transition minus the rounded experimental reading in Ref. [23]. It is not a solver error.

| **State** | $\epsilon_{ns}$ **(meV)** | **Transition (eV)** | **Spectral reading (eV)** | **Difference (meV)** |
|---|---|---|---|---|
| 1$s$ | -318.805202 | 2.091195 | 2.090 | +1.195 |
| 2$s$ | -151.726788 | 2.258273 | 2.250 | +8.273 |
| 3$s$ | -94.510994 | 2.315489 | 2.310 | +5.489 |
| 4$s$ | -65.192873 | 2.344807 | 2.340 | +4.807 |
| 5$s$ | -47.752231 | 2.362248 | 2.360 | +2.248 |

To estimate numerical error independently of experiment, we solve the same radial Hamiltonian by two discretizations. The production finite-volume method represents R($r$) on cell centers. An independent piecewise-linear finite-element calculation represents R($r$) on nodes, with the consistent mass matrix for the measure $r$ dr and 12-point Gaussian integration per element. Its weak form is

$$\frac{\hbar^2}{2\mu}\int ru'v'\,dr + \int rV(r)uv\,dr = E\int ruv\,dr. \tag{S5}$$

Here $u(r)$ and v($r$) are finite-element test and trial functions, prime denotes a radial derivative, and the integrals extend from zero to $R_{max}$. Regularity at zero is the natural boundary condition and v($R_{max}$) = 0. This independent construction does not reuse the production Hamiltonian matrix.

Table S7 compares both methods for the most tightly bound 1$s$ state; the accompanying data include all 1$s$-5$s$ states and residuals.

Table S7. Numerical convergence for the identical screened Hamiltonian. FV and FE denote finite volume and finite element. All energies and differences are in meV; lengths are in nm.

| $R_{max}$ | $\Delta\ r$ | $E_{FV}$ | $E_{FE}$ | $E_{FV}$ - $E_{FE}$ |
|---|---|---|---|---|
| 200 | 0.1 | -319.109444 | -318.756794 | -0.352650 |
| 200 | 0.05 | -318.873769 | -318.772627 | -0.101142 |
| 200 | 0.025 | -318.805202 | -318.776646 | -0.028556 |
| 200 | 0.0125 | -318.785617 | -318.777658 | -0.007959 |
| 300 | 0.025 | -318.805202 | -318.776646 | -0.028556 |

At $\Delta\ r = 0.0125$ nm, the two methods differ by 0.00796 meV for 1$s$ and by at most 0.00251 meV for 2$s$-5$s$. Increasing $R_{max}$ from 200 to 300 nm at $\Delta\ r = 0.025$ nm leaves all five energies unchanged to better than 0.00001 meV. The finite-volume change from $\Delta\ r = 0.025$ to 0.0125 nm is 0.0196 meV for 1$s$, whereas the finite-element change is 0.00101 meV. Thus the 1.2-8.3 meV differences from rounded spectral readings are much larger than the tested radial discretization effects. They compare an approximate fitted effective-mass model with experimental peak estimates; the available information does not separate fitting residuals, rounded input parameters, and spectral-readout uncertainty into unique contributions.

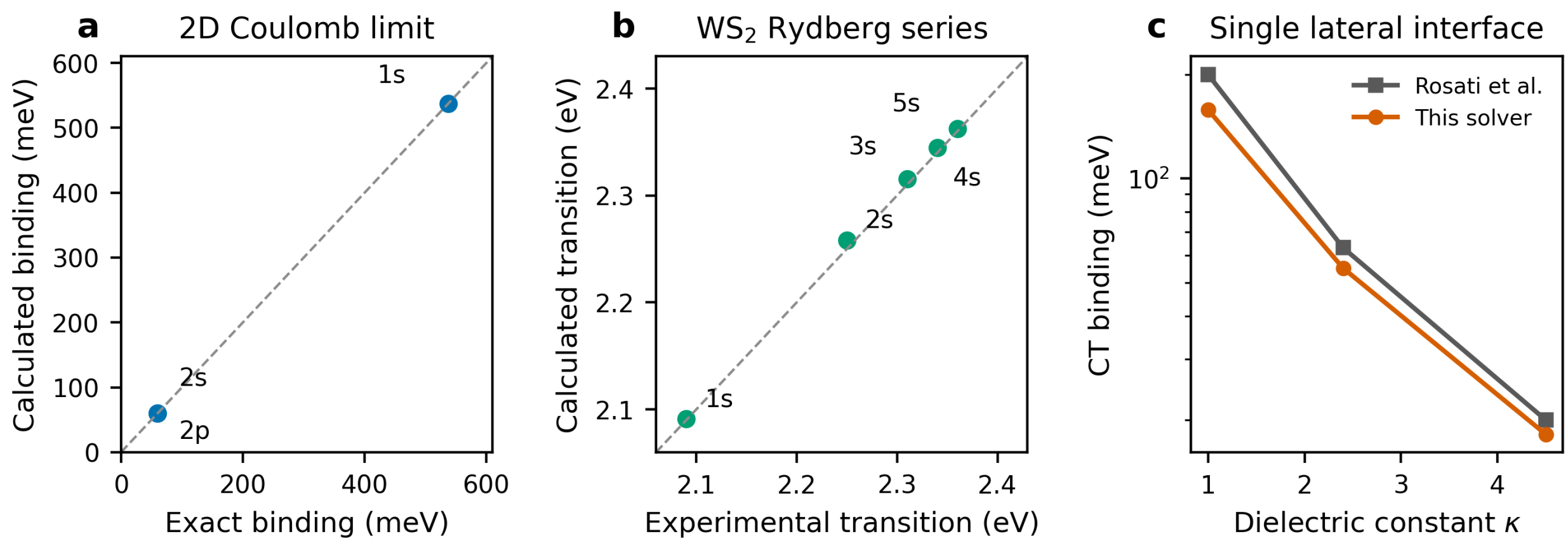

Figure S2: Three distinct reference comparisons. (a) Numerical energies versus the exact pure-Coulomb spectrum. (b) Screened $WS_2$ transitions versus rounded experimental readings from Ref. [23]. ($c$) Present-model single-interface binding versus published calculations in Ref. [9]. Panels (b,$c$) are not comparisons with exact eigenvalues of an identical Hamiltonian. Numerical and model differences are separated in Sec. S4.

### $C$. Single interface comparison and sources of disagreement

We compare a $MoSe_2$-$WSe_2$-like single-interface limit with Ref. [9]. The present calculation uses $m_e = 0.49 m_0$, $m_h = 0.36 m_0$, $\Delta\ E_c = 285$ meV, $\Delta\ E_v = 215$ meV, and $r_0 = 7.5$ nm in Eq. (4). The interface profile is $\tanh(4x/w)$. A central width $L = 80$ nm separates the interfaces; the nearly degenerate lowest pair represents the two independent interface locations. The reported energy is the mean of that pair, and diagonalizing the projected coordinate operator gives the localized separation $d$. This benchmark has different material inputs from the main $WS_2$-$MoS_2$-$WS_2$ calculation.

Table S8. Single-interface comparison with Ref. [9], plotted in Fig. S2($c$). Published values are approximate. These are not established matched-Hamiltonian comparisons.

| $\kappa$ | $w$ **(nm)** | $E_b$ **calculated (meV)** | $E_b$ **reference (meV)** | $d$ **calculated (nm)** | $d$ **reference (nm)** |
|---|---|---|---|---|---|
| 1.0 | 2.4 | 158.05 | 200 | 2.48 | 1 |
| 2.4 | 2.4 | 54.99 | 63 | 4.75 | 5 |
| 4.5 | 2.4 | 18.18 | 20 | 8.90 | 9.6 |
| 4.5 | 12.0 | 7.76 | 7* | 26.4 | 27 |

*The approximately 7 meV wide-interface value quoted in Ref. [9] is for $\Delta\ E_{\mathrm{v}} = 165$ meV, rather than the 215 meV used here. This row illustrates the width trend and is not a matched-offset benchmark.

For $w = 2.4$ nm, the present binding energies are lower than the cited approximate values by about 21%, 13%, and 9% at $\kappa = 1$, 2.4, and 4.5. These differences are not described as sub-percent numerical agreement. Matching the Hamiltonian, parameters, approximations, and energy reference should reproduce a published numerical result within discretization and reporting errors. That full matching has not been established here. In particular, Ref. [9] uses a generalized thin-film screened interaction and a local relative-state/center-of-mass separation, whereas the present calculation uses Eq. (4) with the stated phenomenological length and retains the coupled $X$ and relative motion. The published main text and SI do not provide a tabulated potential kernel and eigenvalue data sufficient to verify a point-by-point match to our implementation.

To quantify errors within our own model, we repeat the $\kappa = 4.5$, $w = 2.4$ nm case while independently enlarging the $X$ domain, halving its spacing, and enlarging the relative basis. Table S9 also compares the coupled solution with a deliberately adiabatic calculation using exactly the same local Hamiltonian matrices. For each $X$, the latter retains the lowest local relative energy $U_0(X)$, then solves $[-\hbar^2\partial_X^2/(2M) + U_0(X)]\chi = E\ \chi$. It omits all $X$ derivatives of the local relative eigenfunction. Thus it is a controlled test of the solution approximation, not a claim to reproduce the complete model of Ref. [9].

Table S9. Numerical and adiabatic diagnostics at fixed single-interface model parameters. $E_b$ is measured from the consistently computed electron-strip threshold. The last column is $E_b$(adiabatic) - $E_b$(coupled). All energies are in meV and lengths in nm.

| **Diagnostic** | $h$ **(nm)** | $E_b$ **coupled** | $E_b$ **adiabatic** | **Difference** |
|---|---|---|---|---|
| Reference grid | 0.4 | 18.17984 | 23.01801 | 4.83817 |
| $X$ half-range 80 nm | 0.4 | 18.17984 | 23.01801 | 4.83817 |
| $X$ spacing 0.2 nm | 0.2 | 18.13320 | 22.99750 | 4.86430 |
| Larger relative basis and quadrature | 0.4 | 18.20003 | 23.05380 | 4.85377 |

Increasing the $X$ half-range from 60 to 80 nm at fixed spacing 0.4 nm has a negligible effect. Reducing that spacing to 0.2 nm changes the coupled binding by -0.04664 meV (0.257%), while enlarging the radial range, angular basis, and quadrature changes it by +0.02019 meV (0.111%). The zero-field electron-strip threshold is about 0.1192 meV; using the zero threshold of an ideal isolated interface would reduce each listed binding by that amount. These identified numerical and reference-energy effects are smaller than the approximately 1.82 meV difference from 20 meV in Table S8. In contrast, omitting the derivatives of the local relative eigenfunction increases binding by 4.84-4.86 meV. This shows that the solution approximation can matter on a larger scale than the observed residual, but it does not assign the entire literature discrepancy to that approximation: the screening kernel and remaining input conventions still require matching. The $\kappa = 1$ and 2.4 discrepancies have not been decomposed by this diagnostic.

Figure S3 displays the original dielectric and width comparisons and the relative-basis refinement. Its panel ($d$) tests relative-basis changes only, not all numerical errors; Table S9 supplies the additional $X$-grid checks.

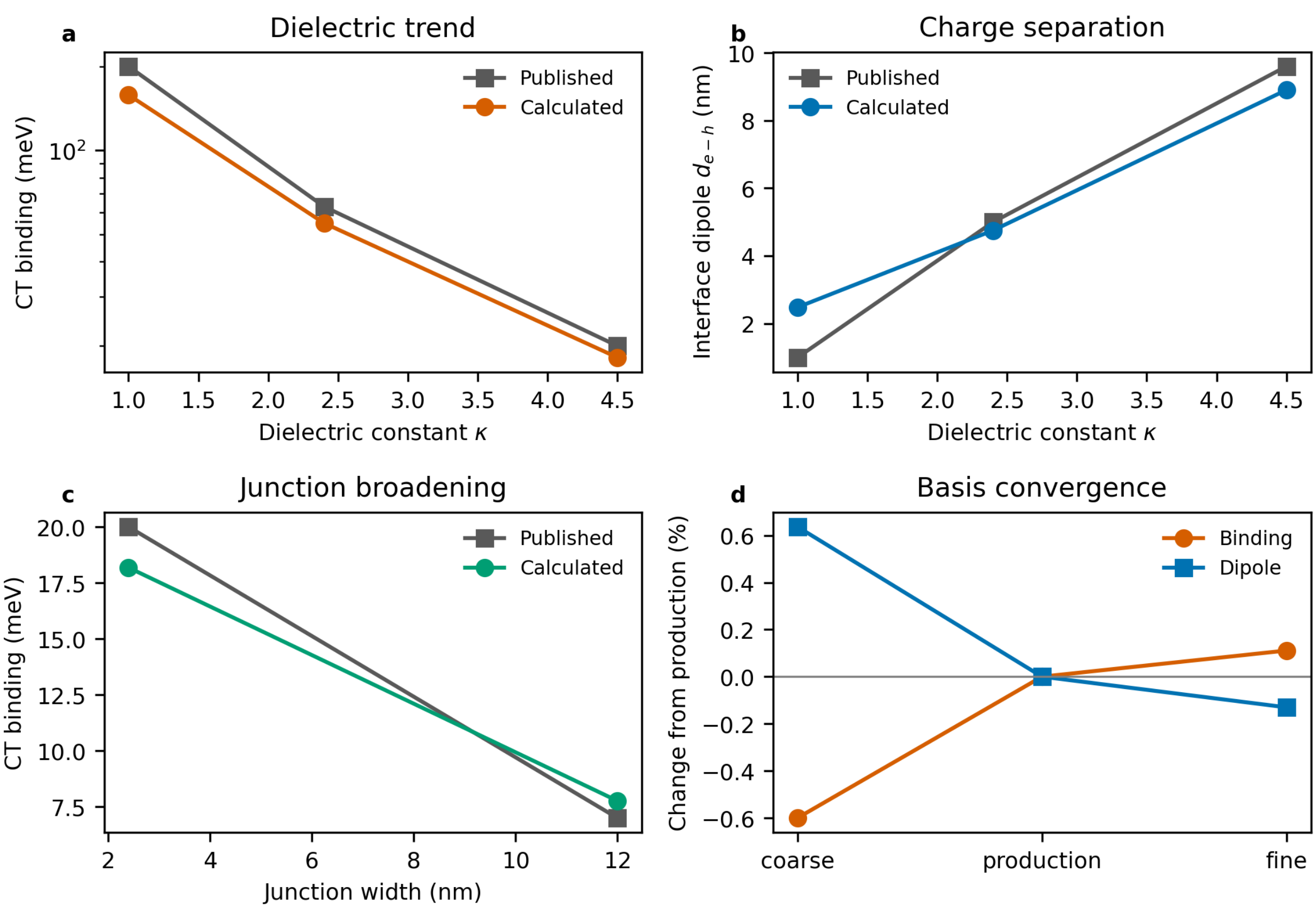

Figure S3: Single-interface model comparison. Dielectric dependence of binding and separation, interface-width dependence, and relative-basis refinement. Published points are approximate values from Ref. [9]. This separate model comparison is not a fit of the double-interface parameter set or a complete error bound; see Table S9 for additional numerical diagnostics.

## S5. Field-induced mixing with higher states

Let $\hat{P}$ project onto the two zero-field CT1 eigenstates, one even and one odd under left-right reflection. The branch labels even-derived and odd-derived refer to these zero-field origins, not finite-field parity. The restricted Hamiltonian is $\hat{P}\ \hat{H}_{\text{red}}\ \hat{P}$, with the same applied $F$ as in the full calculation. It retains the zero-field splitting and the electric matrix element within CT1. The full solution permits mixing with all other spatial states. The weight outside CT1 is one minus the sum of squared overlaps with the two zero-field vectors; it is not the total non-1$s$ basis content of the interface exciton.

Table S10. CT1 projection test in the 36-nm representation. Additional energy shifts are full minus projected values in meV; W is the probability outside CT1.

| $F/F_c$ | **Branch** | **W (%)** | **Energy difference** |
|---|---|---|---|
| 1 | even-derived | 0.02809 | -0.015289 |
| 1 | odd-derived | 0.05992 | -0.029987 |
| 3 | even-derived | 0.40210 | -0.188461 |
| 3 | odd-derived | 0.47643 | -0.229768 |

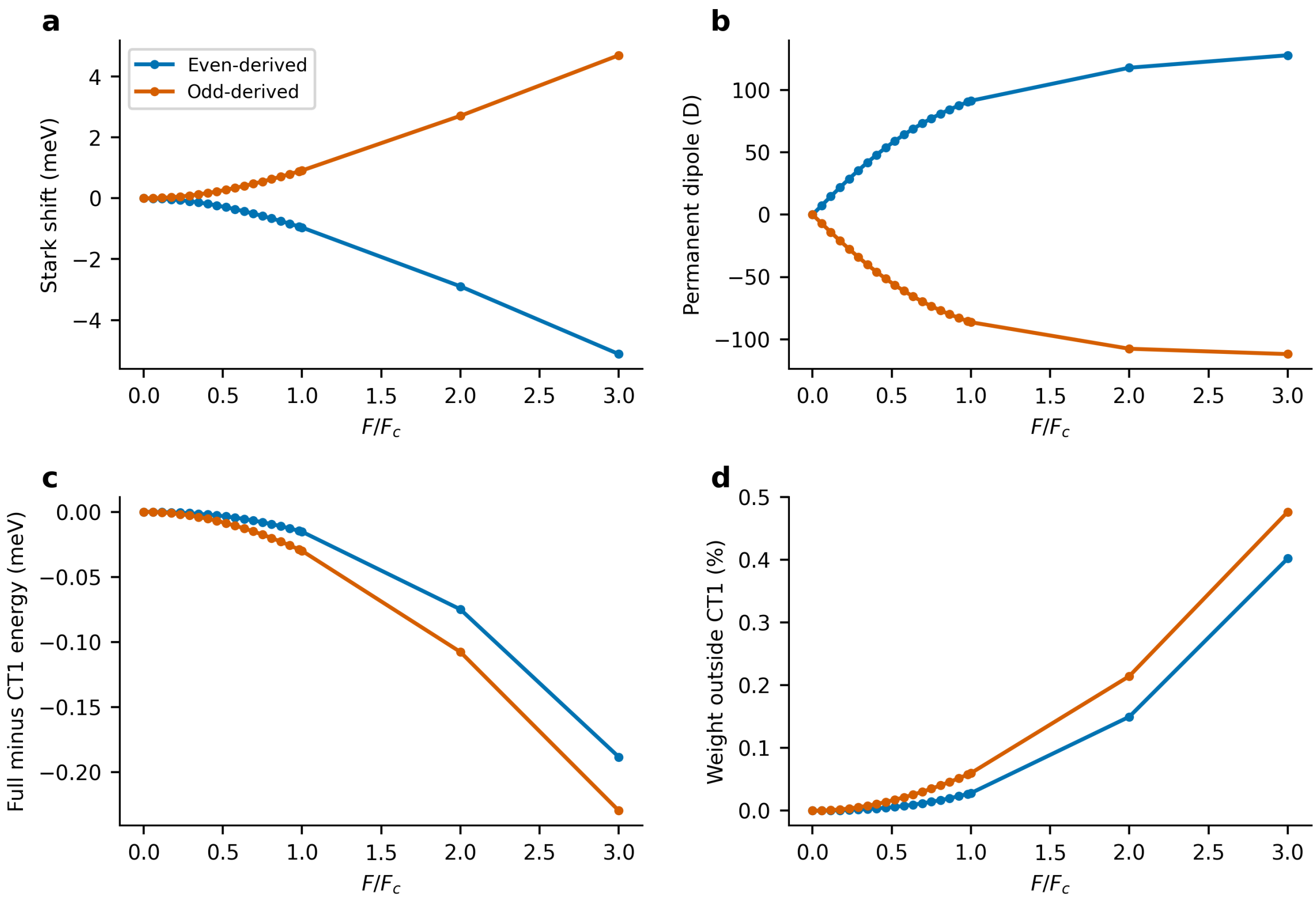

Figure S4: Full spatial calculation at $L = 1.5$ nm. (a) Stark shifts relative to each zero-field branch. (b) Permanent dipoles. ($c$) Energy corrections beyond CT1. ($d$) Probability outside CT1. Blue and red denote the even-derived and odd-derived branches. Numerical domains are specified in Sec. S3.

## S6. Parameter sensitivity and width sweep

Every sensitivity case uses the 204-function production representation at $L = 1.5$ nm, except the explicitly labeled strip-width endpoints. The separated-carrier energy is recomputed for each parameter set. Figure S5 normalizes each result by the unchanged Table S1 inputs.

Table S11. One-at-a-time sensitivities. Energies are in meV, lengths in nm, and fields in V/$\mu$m.

| **Case** | $E_b$ | **2**$J$ | $d$ | $F_c$ |
|---|---|---|---|---|
| Table S1 | 104.280 | 4.57084 | 2.6300 | 0.86898 |
| $\Delta$ $E_c$ 488 | 105.196 | 4.59640 | 2.5992 | 0.88421 |
| $\Delta$ $E_c$ 732 | 103.606 | 4.54353 | 2.6477 | 0.85802 |
| $\Delta$ $E_v$ 360 | 109.303 | 7.88023 | 2.4552 | 1.60481 |
| $\Delta$ $E_v$ 540 | 101.144 | 2.82860 | 2.7488 | 0.51452 |
| $w$ 0.7 | 105.395 | 4.69195 | 2.5871 | 0.90681 |
| $w$ 1.3 | 103.094 | 4.44184 | 2.6801 | 0.82868 |
| $\kappa$ 2 | 129.500 | 5.46196 | 2.4790 | 1.10167 |
| $\kappa$ 3 | 85.769 | 3.84849 | 2.7830 | 0.69143 |
| $r_0$ 6 | 117.173 | 6.45767 | 2.4007 | 1.34493 |
| $r_0$ 9 | 94.658 | 3.46722 | 2.8239 | 0.61390 |
| mass 0.9 | 100.624 | 5.14783 | 2.7398 | 0.93945 |
| mass 1.1 | 107.573 | 4.07525 | 2.5383 | 0.80275 |
| $L$ 1.4 | 106.481 | 5.60492 | 2.5454 | 1.10100 |
| $L$ 4 | 75.484 | 0.03171 | 4.3130 | 0.00368 |

The width sweep in Fig. 2(b-$d$) uses the same representation at every $L$. The $L = 1.5$ nm point is recomputed directly, not interpolated. The numerical-domain checks in Sec. S3 apply at the representative width; endpoint values in the width plot use the common production discretization.

The sensitivity scan varies each band offset by 20%, $w$ from 0.7 to 1.3 nm, $\kappa$ from 2.0 to 3.0, $r_0$ from 6 to 9 nm, and both masses together by 10%, keeping the other inputs fixed. Across these variations, the even-state binding spans 85.8-129.5 meV and $F_c$ spans 0.515-1.605 V/$\mu$m. Increasing the hole barrier from 360 to 540 meV lowers $F_c$ from about 1.60 to 0.51 V/$\mu$m while changing the binding less strongly.

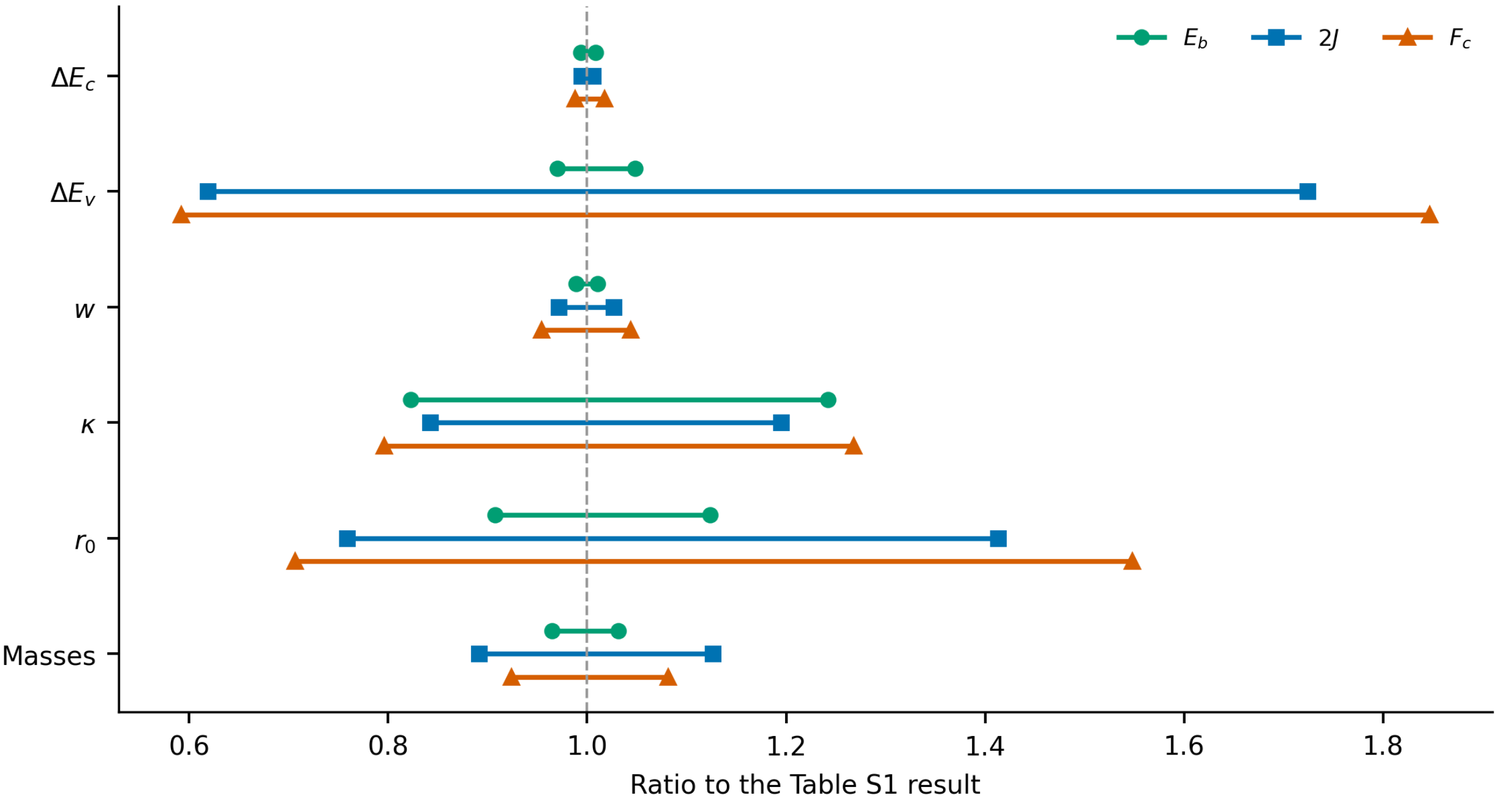

Figure S5: Sensitivity at $L = 1.5$ nm, normalized to the Table S1 inputs. Green circles, blue squares, and red triangles indicate binding energy, splitting $2J$, and crossover field, respectively. The line endpoints are individual calculations, not uncertainty bounds. Numerical values are given in Table S11 and the plotting-data archive.

## S7. Static detuning and optical assumptions

The sign convention is $\Delta_0 = E_\mathrm{L}$ - $E_\mathrm{R}$ and $\hat{\rho}_z = |\mathrm{R}\rangle\langle\mathrm{R}|$ - $|\mathrm{L}\rangle\langle\mathrm{L}|$. Hence the bias term is -$(edF + \Delta_0/2)\hat{\rho}_z$, canceled at $F_\mathrm{sym}$ = -$\Delta_0/(2ed)$. The minimum gap stays $2J$ when $J$ and $d$ are held fixed. A lower-state hybridization expectation of at least $1/\sqrt{2}$ corresponds to $|edF + \Delta_0/2|$ no larger than $J$.

The ideal optical formulas assume identical complex optical amplitudes for the two localized configurations, in the same phase convention as the Hamiltonian. More generally, the transition amplitude is $C_\mathrm{L}a_\mathrm{L} + C_\mathrm{R}a_\mathrm{R}$. Equation (10) follows from the local interband dipole operator: at zero photon momentum, integration over a normalized longitudinal plane wave gives $\sqrt{L_y}$, and the transverse integral is evaluated at zero electron-hole separation. If the Bloch dipole varies across the interface, $d_\mathrm{cv}$ must instead remain inside the $X$ integral. The optical strength is the squared amplitude, not the integral of its local square.

Specifically, integrating $|\Psi_a(X,0)|^2$ over $X$ gives $C_a$, the integrated local electron-hole coincidence density, sometimes called the contact. This local coincidence density can be finite for both parity states. In contrast, the odd state's $S_a$ vanishes because the amplitude is odd under $X$ reflection. Figure 5 displays transition energies and relative dipole strengths, rather than the local coincidence density.

The spectral maps use a Lorentzian half-width $\gamma$ = 0.25 meV (full width 0.50 meV) and the sum of $f_a\ \gamma/[\pi((E-E_a)^2+\gamma^2)]$. Here $f_a$ is the computed $S_a/S_\mathrm{CT1,even}(0)$. The maps show discrete exciton contributions to absorption; computing reflectance contrast additionally requires the substrate optical stack. The field-dependent cancellation and activation of the odd-derived branch remain the optical signatures of interest.

## S8. Data and code archive

The accompanying archive includes the main and Supplemental Material Word files, directly compiled REVTeX/article PDFs, PNG figures, and CSV numerical plotting data. The current figure index identifies the data for every figure panel. The representative_$L1p5$ directory contains the eigensolutions, carrier occupations, projection checks, and sensitivity calculations for $L$ = 1.5 nm. Data for the width dependence and reference-system checks are identified separately. Solver sources and tests are retained under source_snapshot.

## S9. Excited states and numerical domains

The cosine angular basis spans states even under $y$ reflection. It includes angular and radial mixing and is not restricted to a hydrogenic $s$ series. Table S12 gives the CT1 and CT2 parity pairs used in the main article.

Table S12. Zero-field pairs at $L$ = 1.5 nm; energies are relative to the electron-strip threshold. Binding and splitting are in meV.

| **Domain** | **Pair** | $E_{b,even}$ | $E_{b,odd}$ | **2**$J$ |
|---|---|---|---|---|
| 36 | 1 | 104.2941 | 99.7313 | 4.56274 |
| 36 | 2 | 54.8555 | 52.5963 | 2.25923 |
| 44 | 1 | 104.2991 | 99.7363 | 4.56285 |
| 44 | 2 | 54.9629 | 52.7256 | 2.23731 |
| 52 | 1 | 104.3045 | 99.7419 | 4.56266 |
| 52 | 2 | 54.9846 | 52.7530 | 2.23162 |

The CT1 and CT2 levels are interpreted as localized excitons over the field window in Fig. 5, where the domain comparison in Sec. S11 supports their energies and optical strengths.

For Fig. 5, the low-field CT1 and CT2 energies and strengths are calculated in the 36-nm representation at steps of 0.05 V/$\mu$m through 0.85 V/$\mu$m. Linear interpolation in field is used only to display the optical map. The 44-nm check at 0.85 V/$\mu$m is tabulated in Sec. S11. A Lorentzian half-width of 0.25 meV broadens the four discrete levels for display; no continuum background is added.

## S10. Electric-field quantum metric and strip width

For the normalized nondegenerate lower eigenstate $|u(F;L)\rangle$, we define the quantum metric with respect to the physical electric field $F$ [24]:

$$g_{FF} = \mathrm{Re}\langle \partial_F u|(\hat{I} - |u\rangle\langle u|)|\partial_F u\rangle, \tag{S6}$$

The strip width $L$ labels separate structures at which the field derivative is evaluated. Thus this is $g_{FF}(F;L)$, not the width-derivative metric $g_{LL}$, nor a Bloch-momentum metric. For $\hat{H}_{\mathrm{CT}} = E_0\hat{I}$ - $J$ $\hat{\rho}_x$ - $q$ $\hat{\rho}_z$, $q = edF$ at equivalent interfaces, the two-state result is

$$g_{FF}(F;L) = \frac{(ed)^2J^2}{4(J^2+q^2)^2}, \qquad g_{FF}(0;L) = \frac{1}{4F_{\mathrm{c}}^2}. \tag{S7}$$

At zero field, $g_{FF} = (ed)^2/(4J^2) = 1/(4F_{\mathrm{c}}^2)$. With $F$ in V/$\mu$m, the metric is reported in $(\mu\mathrm{m/V})^2$. Equivalently, the full eigenstate spectral expression is the sum over excited states of $|\langle n|ex|0\rangle|^2/(E_n - E_0)^2$. The CT1 odd partner gives the two-state contribution, while higher states add positive terms.

We calculate the full metric independently using the fidelity between the lowest spatial eigenstates at -$\delta$ $F$ and +$\delta$ $F$: $g_{FF}(0)$ = [1 - $|\langle u(-\delta F)|u(+\delta F)\rangle|^2$]/(4 $\delta$ $F^2$) in the small-step limit. Left-right reflection generates the negative-field state from the positive-field solution. We use $\delta$ $F$ = 0.002$F_{\mathrm{c}}$ and 0.001$F_{\mathrm{c}}$ to check the step size at every width. The numerical table also gives the positive spectral sum over the lowest eight states, which is a partial sum rather than an independent complete-basis result.

Across $L$ = 1.4-4.0 nm, the full-wavefunction metric increases from 0.2064 to 18502.4 $(\mu\mathrm{m/V})^2$. At $L$ = 2 nm it is 3.30030 $(\mu\mathrm{m/V})^2$. The largest relative finite-difference step change is 0.00030%. The full result exceeds the CT1 expression by at most 0.1022% in this width range, showing that the same lowest-pair hybridization controls the zero-field geometric response. At the widest widths the difference is comparable to the small finite-step truncation error; the archived partial spectral sums provide a positive check on the higher-state contribution.

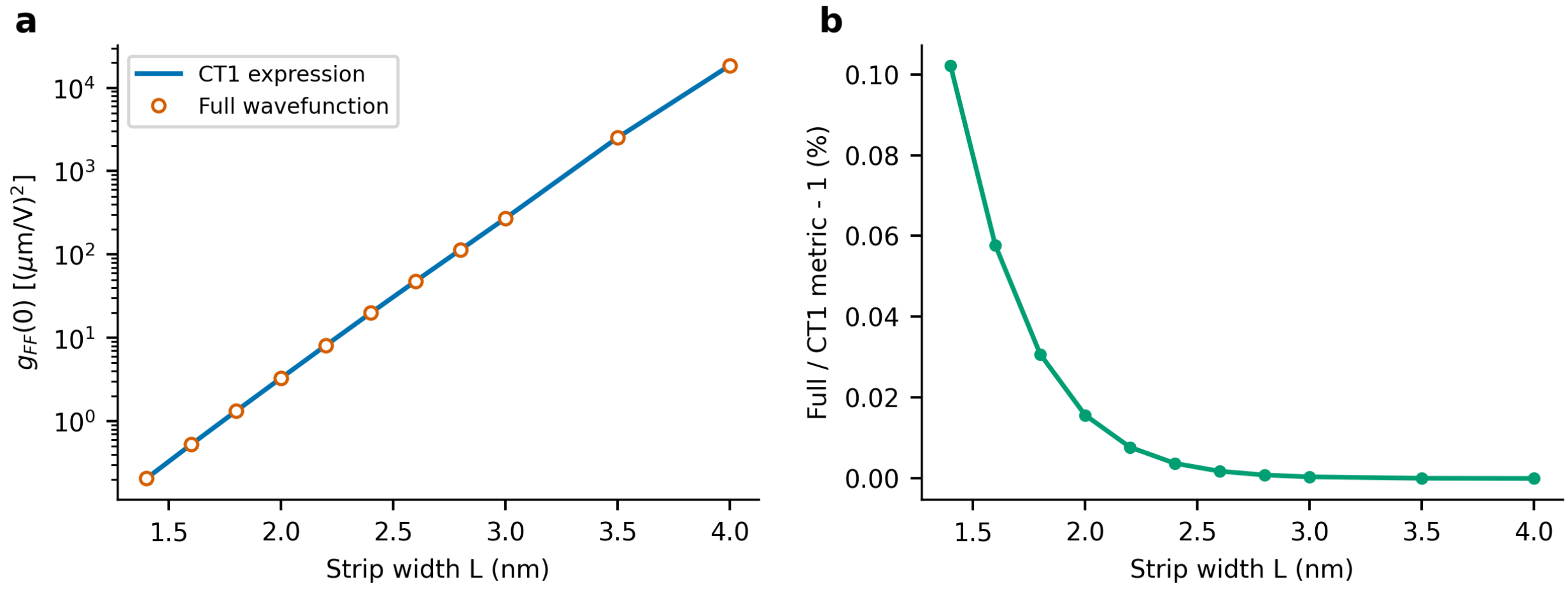

Figure S6: Quantum metric of the lower CT state at zero electric field. (a) Width dependence of $g_{FF}$ from the two-state expression and from full-wavefunction finite differences. (b) Relative difference between the full result and the two-state contribution. The same 204-function basis is used at each $L$, with $w$ = 1 nm. The finite-difference step check and all numerical values are supplied in the accompanying CSV table.

The larger metric at wider strips expresses a more rapid change of the state under a fixed physical field increment, resulting from the decreasing avoided-crossing gap. It is not a coherence time or gate fidelity. Rescaling the field to $u = F/F_c$ gives $g_{uu}(0) = 1/4$ in the two-state model, making explicit that the width dependence in panel (a) measures the physical control scale. At the exact $J = 0$ degeneracy, the nondegenerate-state metric used here is not defined.

## S11. Pair-dependent coupling and optical response

Each zero-field pair has $J_n = (E_{odd,n}$ - $E_{even,n})/2$, $d_n = |\langle \text{even}|x|\text{odd}\rangle|$ within that pair, and $F_{c,n} = J_n/(ed_n)$.

Table S13. Pair-dependent scales for $L = 1.5$ nm.

| **Domain** | **Pair** | $J$ **(meV)** | $d$ **(nm)** | $F_c$ **(V/$\mu$m)** |
|---|---|---|---|---|
| 36 | 1 | 2.28137 | 2.6318 | 0.86686 |
| 36 | 2 | 1.12961 | 5.9944 | 0.18844 |
| 44 | 1 | 2.28142 | 2.6317 | 0.86691 |
| 44 | 2 | 1.11865 | 5.9913 | 0.18671 |
| 52 | 1 | 2.28133 | 2.6315 | 0.86692 |
| 52 | 2 | 1.11581 | 5.9906 | 0.18626 |

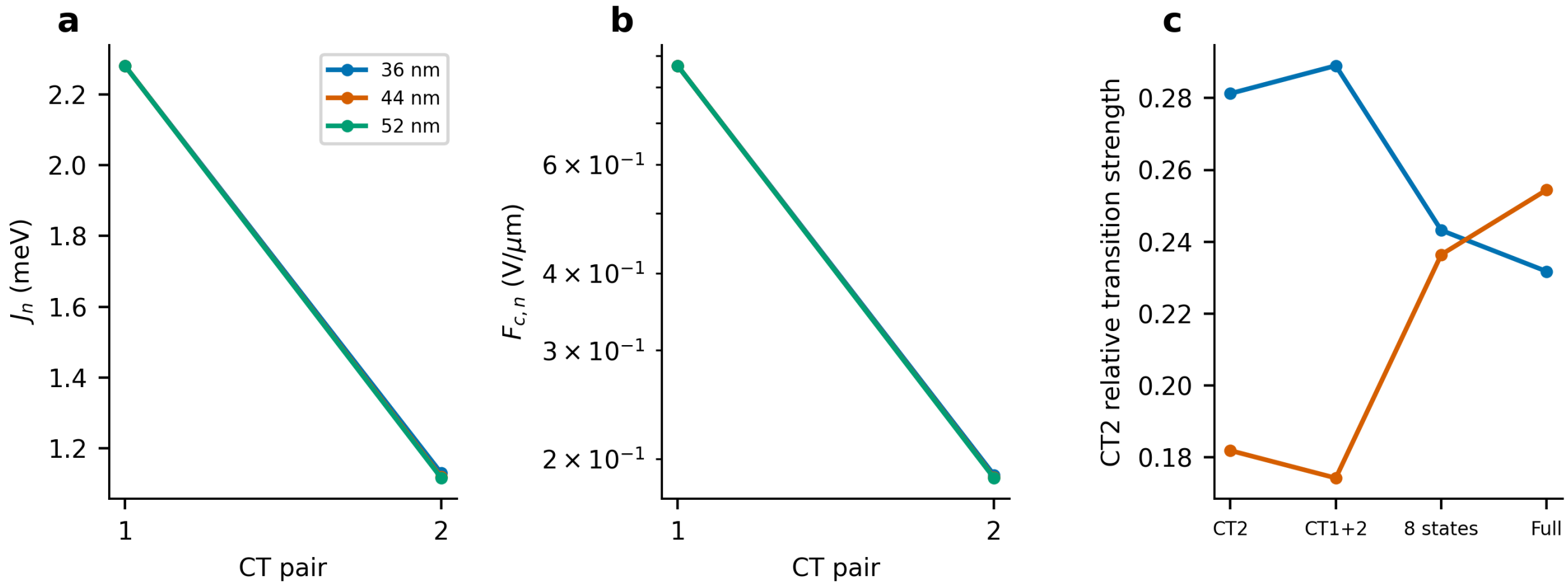

Figure S7: CT1 and CT2 response at $L = 1.5$ nm. (a,b) Couplings and electric scales for three numerical representations. ($c$) CT2 transition strengths at 0.85 V/$\mu$m, obtained from an isolated CT2 pair, CT1 plus CT2, eight zero-field states, and full spatial diagonalization in the 44-nm representation. Blue and vermilion distinguish the CT2 even-derived and odd-derived states.

Table S14. Full CT2 optical strengths and outer probabilities at 0.85 V/$\mu$m; $P_{out}$ integrates separations larger than 0.8R.

| **Domain** | **Branch** | **Relative strength** | $P_{out}$ |
|---|---|---|---|
| 36 | even-derived | 0.23437 | 7.650$e$-07 |
| 36 | odd-derived | 0.25529 | 1.510$e$-06 |
| 44 | even-derived | 0.23174 | 5.831$e$-07 |
| 44 | odd-derived | 0.25444 | 7.901$e$-07 |

## S12. Spatial extent and CT2 coupling

For CT1 the average hole probability in the central $MoS_2$ region is 0.029987; for CT2 it is 0.013450. These averages equal the occupation in either localized combination, because the central-region operator is reflection even. The larger CT2 radius therefore accompanies lower central-barrier occupation. The coupling remains a Hamiltonian matrix element and cannot be inferred from radius alone. The regional probabilities are given in Table S2, and the phase-sensitive wavefunctions in Fig. 3.

## S13. Center-of-mass density and reflection

The transverse center-of-mass density is

$$P_a(X) = \int dx\, dy\, |\Psi_a(X,x,y)|^2 = \sum_j |\chi_{aj}(X)|^2, \qquad \int dX\, P_a(X) = 1. \tag{S8}$$

It integrates over both relative coordinates. A signed $y = 0$ wavefunction section, as in Fig. 3, is different: it retains the conditional dependence on both carrier positions. Reflection through $x = 0$ maps ($x_e$, $x_h$) to (-$x_e$, -$x_h$) simultaneously. A cut as a function of $x_h$ with nonzero $x_e$ held fixed need not be symmetric. The even and odd states obey this simultaneous reflection with signs +1 and -1. The displayed maps are checked against that relation.

The zero-field envelope symmetries are $P_x$:($X$,$x$,$y$) to (-$X$,-$x$,$y$) and $P_y$:($X$,$x$,$y$) to ($X$,$x$,-$y$). Their product is the twofold rotation about the layer normal and supplies no independent label. A transverse electric field breaks $P_x$ but preserves $P_y$. The sine sector is odd in $y$ and is dark in the local optical model. Within the computed even-in-$y$ sector, odd-in-$x$ states are dark at zero field and acquire optical strength through field mixing. These are envelope symmetries, not claims about atomistic inversion symmetry of a 1$H$ monolayer.

## S14. Internal polarization and the two-state asymptote

In a CT1-only projection the localized configurations retain their zero-field dipoles plus or minus $ed$. Increasing $F$ changes their weights, so the projected permanent dipole approaches these values. Full diagonalization also changes their internal charge distributions. Figure S4(b) includes this effect. The probability outside CT1 is not by itself an error estimate for a dipole or optical strength: observables can receive interference contributions linear in a small added amplitude. The CT1-projected uniform-field zero-field polarizability is $\alpha_0 = (ed)^2/J$, and $\alpha_0 F_c = ed$. For $L = 1.5$ nm the two-state value is approximately 145.8 D/(V/$\mu$m), with localized dipole 126.4 D. The larger $J$ compared with $L = 2$ nm improves energy splitting while lowering this differential response. An independent central finite difference of the full CT1 even-derived energy, using $\delta\ F = 0.001$ V/$\mu$m in the 28-nm production representation, gives a curvature magnitude of 3.0577 meV/(V/$\mu$m)$^2$, or 146.87 D/(V/$\mu$m). The CT1-only value in that same representation is 3.0265 meV/(V/$\mu$m)$^2$, or 145.37 D/(V/$\mu$m); higher-state mixing increases it by approximately 1.03%. The ratios to the vertical benchmark below use CT1-projected responses consistently.

For a quantitative vertical comparison, we use the D5 fit in Supplemental Section 9 of Ref. [12]: opposite dipole lengths 0.659 and 0.740 nm, and off-diagonal coupling 12.0 meV. In our notation $J$ is this coupling, not twice its value. Their mean length is 0.6995 nm. Removing the common linear energy tilt gives a differential response of 1.9585 D/(V/$\mu$m) at the compensation field, approximately 3.00 V/$\mu$m from the fitted zero-field energy difference of 4.2 meV. The unequal dipoles leave a common linear tilt but do not change this curvature.

The archived $L = 2.0$ nm calculation gives $J = 0.832100$ meV, $d = 3.02836$ nm, and $\alpha_0 =$ 529.383 D/(V/$\mu$m), hence a ratio of 270.30 to this vertical reference. The $L = 1.5$ nm result is approximately 74.5 times the same reference. The ratio compares lateral zero field with vertical compensation, where the hybridization curvature is largest, using the local electric-field convention of each model. It is not a ratio of response per applied gate voltage. Numerical inputs and the comparison are provided in plot_data/Stark_curvature_comparison.csv; the $L = 2.0$ nm source metadata are retained separately from the representative 1.5-nm results.

## S15. Internal radius and homogeneous two-dimensional reference

The relative density and radial separation distribution are

$$\begin{gathered} p_a(x,y) = \int dX\,|\Psi_a(X,x,y)|^2, \\ P_{r,a}(r) = r\int_0^{2\pi} d\theta\, p_a(r\cos\theta, r\sin\theta). \end{gathered} \tag{S9}$$

The radial factor $r$ is the polar area element; the probability integrates over dr to one. The effective radius is

$$r_{\mathrm{rms},a} = \sqrt{\int dr\, r^2 P_{r,a}(r)} = \sqrt{\langle x^2 + y^2\rangle_a}. \tag{S10}$$

Removing the band-offset potentials while retaining the same masses and screening gives a homogeneous 2D 1$s$ state with binding 198.06 meV and rms radius 1.7537 nm. This isolates the geometric effect and is not a separately calibrated pristine-material prediction.

Table S15. Internal rms radii at $L = 1.5$ nm in the 52-nm representation.

| **State** | $r_{\mathrm{rms}}$ **(nm)** |
|---|---|
| CT1 even | 3.39743 |
| CT1 odd | 3.63501 |
| CT2 even | 7.31528 |
| CT2 odd | 7.69735 |

For either localized combination, the squared radius decomposes into the positive mean-separation length $d$ and variances. The signed expectation value of $x$ is opposite in $|\mathrm{L}\rangle$ and $|\mathrm{R}\rangle$:

$$\begin{gathered} d = |\langle x\rangle|, \quad \sigma_x^2 = \langle x^2\rangle - d^2, \quad \sigma_y^2 = \langle y^2\rangle, \\ r_{\mathrm{rms}}^2 = d^2 + \sigma_x^2 + \sigma_y^2. \end{gathered} \tag{S11}$$

For CT1, $d = 2.63153$ nm, $\sigma_x = 1.36295$ nm, and $\sigma_y = 1.89614$ nm. The radial distributions use exact cosine-channel orthogonality and finite-volume radial inner products. Their normalizations and rms radii are checked against the eigenvector moments.

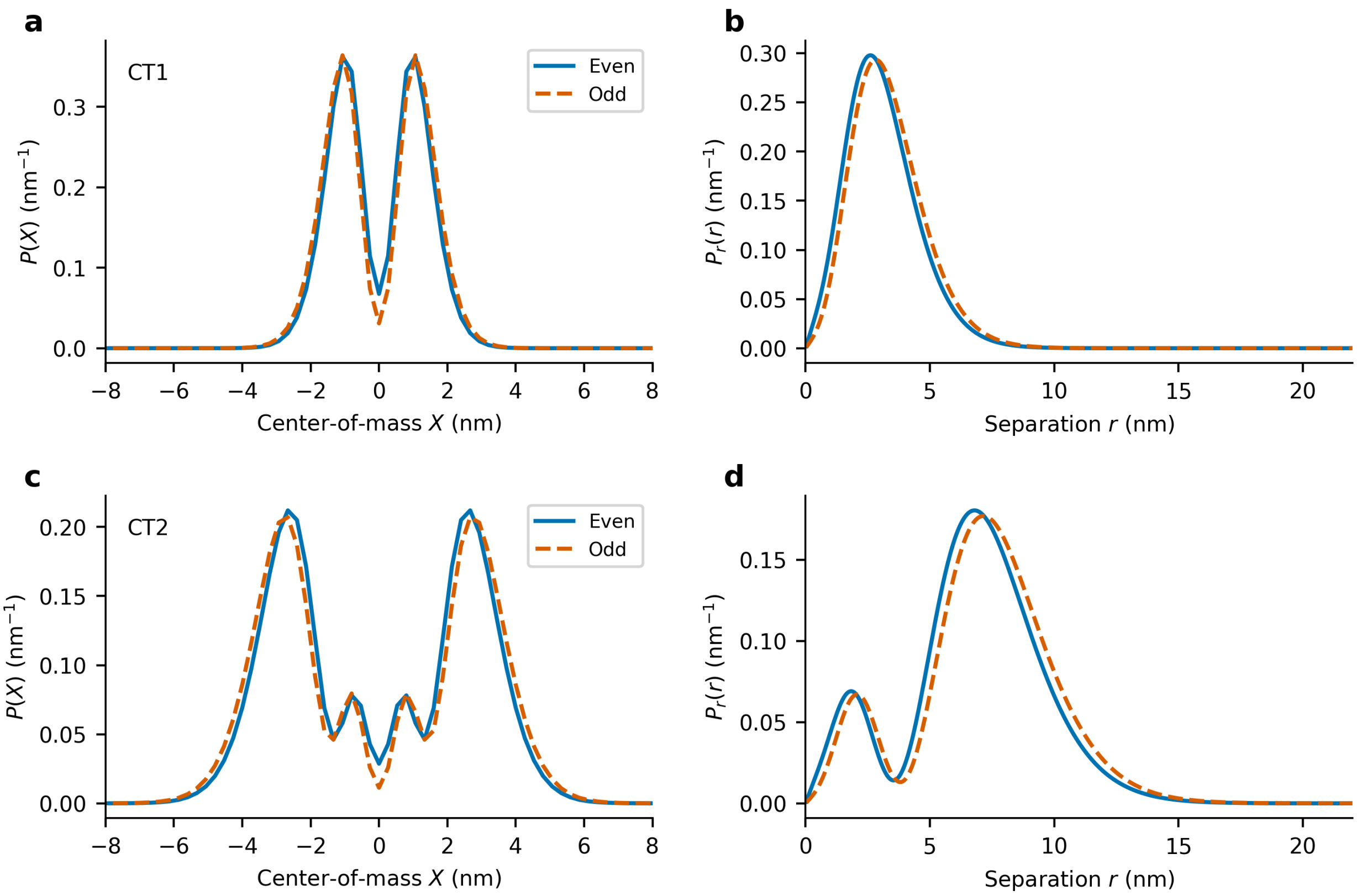

Figure S8: Probability distributions at $L = 1.5$ nm. (a,$c$) Transverse center-of-mass densities and (b,$d$) radial separation distributions. Top and bottom rows show CT1 and CT2; blue solid and orange dashed lines show even and odd states. All curves use the same 52-nm eigenvectors as Fig. 3. Similar probability densities do not imply equal wavefunction phases.

## S16. Electric quadrupole moment of the zero-field states

The electric quadrupole is evaluated from the excess envelope charge of one electron and one hole, with charges -$e$ and +$e$. We use the traceless Cartesian convention

$$Q_{ij} = e\langle 3r_{\mathrm{h},i}r_{\mathrm{h},j} - r_{\mathrm{h}}^2\delta_{ij} - 3r_{\mathrm{e},i}r_{\mathrm{e},j} + r_{\mathrm{e}}^2\delta_{ij}\rangle. \tag{S12}$$

Here $r_{\mathrm{e}}$ and $r_{\mathrm{h}}$ are carrier positions, i and $j$ are Cartesian indices, and $\delta_{ij}$ is the Kronecker $\delta$. The ribbon center sets $x = 0$, and both charges lie at z = 0. A normalized, separable center-of-mass wave packet regularizes the free $y$ motion; its common second moment cancels between the charges. Atomic-scale Bloch charge and environmental polarization are not included.

Both zero-field parity states have zero permanent dipole, but their electron and hole second moments need not coincide. Defining $\Delta_x$ as the expectation value of $x_{\mathrm{h}}^2$ - $x_{\mathrm{e}}^2$, and $\Delta_y$ analogously, gives

$$\begin{aligned} \Delta_x &= -2\langle Xx\rangle + \frac{m_{\mathrm{e}} - m_{\mathrm{h}}}{M}\langle x^2\rangle, \\ \Delta_y &= \frac{m_{\mathrm{e}} - m_{\mathrm{h}}}{M}\langle y^2\rangle. \end{aligned} \tag{S13}$$

The $X$-$x$ correlation must be retained. Longitudinal reflection makes the off-diagonal moments vanish; the diagonal components are

$$\begin{aligned} Q_{xx} &= e(2\Delta_x - \Delta_y), \\ Q_{yy} &= e(2\Delta_y - \Delta_x), \\ Q_{zz} &= -e(\Delta_x + \Delta_y). \end{aligned} \tag{S14}$$

Table S16 gives the $L = 1.5$ nm, $w = 1$ nm charge quadrupoles. The 44- and 52-nm calculations use 538 and 766 relative basis functions. Both CT1 parity states have zero permanent dipole and a nonzero quadrupole tensor.

Table S16. CT1 quadrupole tensor; components are divided by $e$ and expressed in $\mathrm{nm}^2$.

| **Domain** | **State** | $Q_{xx}/e$ | $Q_{yy}/e$ | $Q_{zz}/e$ |
|---|---|---|---|---|
| 44 | even | 15.26966 | -6.95432 | -8.31535 |
| 44 | odd | 17.88423 | -8.19327 | -9.69095 |
| 52 | even | 15.26834 | -6.95360 | -8.31473 |
| 52 | odd | 17.88244 | -8.19230 | -9.69014 |

The trace is zero by construction, and the angular matrix elements and coordinate-transform identities are tested independently. The $X$-$x$ correlation is evaluated before forming the tensor.

A central electron and one hole shared equally between $x$ = -$d$ and +$d$ have zero dipole but $Q_{xx}$ = 2$e$ $d^2$ and $Q_{yy} = Q_{zz}$ = -$e$ $d^2$. Thus the two $h$ symbols in Fig. 1($c$) represent one hole, not two. An incoherent equal-weight mixture can also have a quadrupole; coherent hybridization is characterized here by the even-odd splitting and phase-sensitive optical amplitudes.

The permanent quadrupole couples to an electric-field gradient. In contrast, the quadratic Stark shift in a uniform field arises from dipole-induced mixing and the associated polarizability. These are distinct properties of the coupled CT states.